\documentclass[aps,prb,altaffillsymbol,amsmath,amssymb,amsfonts,superscriptaddress,reprint]{revtex4-1}
\usepackage{graphicx}
\usepackage{tabularx}
\usepackage{color}
\usepackage{bm}
\usepackage{latexsym}
\usepackage{dsfont}
\usepackage{epstopdf}
\usepackage{CJK}
\usepackage{gensymb}
\usepackage{multirow}
\usepackage[colorlinks=true]{hyperref}

\newcommand{\curho}{CuRh$_2$O$_4$}
\newcommand{\corho}{CoRh$_2$O$_4$}

\def\ie{{i.e.}}
\def\eg{{e.g.}}
\newcommand{\CC}{C\nolinebreak\hspace{-.05em}\raisebox{.4ex}{\tiny\bf +}\nolinebreak\hspace{-.10em}\raisebox{.4ex}{\tiny\bf +}}

\begin{document}
\title{Spin order and dynamics in the diamond-lattice Heisenberg antiferromagnets CuRh$_2$O$_4$ and CoRh$_2$O$_4$}
\author{L.\,Ge}
\affiliation{School of Physics, Georgia Institute of Technology, Atlanta, GA 30332, USA}
\author{J.\,Flynn}
\affiliation{Department of Chemistry, Oregon State University, Corvallis, OR 97331, USA}
\author{J.A.M.\,Paddison}
\altaffiliation{Present Address: Churchill College, University of Cambridge, Storey’s Way, Cambridge CB3 0DS, UK}
\affiliation{School of Physics, Georgia Institute of Technology, Atlanta, GA 30332, USA}
\author{M.B.\,Stone}
\affiliation{Quantum Condensed Matter Division, Oak Ridge National Laboratory, Oak Ridge, TN 37831, USA}
\author{S.\,Calder}
\affiliation{Quantum Condensed Matter Division, Oak Ridge National Laboratory, Oak Ridge, TN 37831, USA}
\author{M.A.\,Subramanian}
\affiliation{Department of Chemistry, Oregon State University, Corvallis, OR 97331, USA}
\author{A.P.\,Ramirez}
\affiliation{Department of Physics,University of California, Santa Cruz, CA 95064, USA}
\author{M.\,Mourigal}
\affiliation{School of Physics, Georgia Institute of Technology, Atlanta, GA 30332, USA}

\date{\today}
\begin{abstract} 
	Antiferromagnetic insulators on the diamond lattice are candidate materials to host exotic magnetic phenomena ranging from spin-orbital entanglement to degenerate spiral ground-states and topological paramagnetism. Compared to other three-dimensional networks of magnetic ions, such as the geometrically frustrated pyrochlore lattice, the investigation of diamond-lattice magnetism in real materials is less mature. In this work, we characterize the magnetic properties of model A-site spinels CoRh$_2$O$_4$ (cobalt rhodite) and CuRh$_2$O$_4$ (copper rhodite) by means of thermo-magnetic and neutron scattering measurements and perform group theory analysis, Rietveld refinement, mean-field theory, and spin wave theory calculations to analyze the experimental results. Our investigation reveals that cubic CoRh$_2$O$_4$  is a canonical $S\!=\!3/2$ diamond-lattice Heisenberg antiferromagnet with a nearest neighbor exchange $J\!=\!0.63$~meV and a N\'eel ordered ground-state below a temperature of 25 K. In tetragonally distorted CuRh$_2$O$_4$, competiting exchange interactions between up to third nearest-neighbor spins lead to the development of an incommensurate spin helix at 24 K with a magnetic propagation vector $\mathbf{k}_{\rm m}\!=\!(0,0,0.79)$. Strong reduction of the ordered moment is observed for the $S\!=\!1/2$ spins in CuRh$_2$O$_4$ and captured by our $1/S$ corrections to the staggered magnetization. Our work identifies CoRh$_2$O$_4$ and CuRh$_2$O$_4$ as reference materials to guide future work searching for exotic quantum behavior in diamond-lattice antiferromagnets.
\end{abstract}
\maketitle

%================  ================  ================  ================
\section{Introduction}
%================  ================  ================  ================

Antiferromagnetic insulators often host novel forms of magnetic matter dominated by strong quantum fluctuations. Low dimensionality,~\cite{Affleck_1989,Mikeska_2004,Lake_2005,Coldea_2010} geometrical frustration,~\cite{Ramirez_1994,Lee_2008,Han_2012,Savary_2016} spin-orbit coupling~\cite{Jackeli_2009,Banerjee_2016} or topology~\cite{Chisnell_2015,Hirschberger_2015,Chernyshev_2016} are known ingredients to suppress classical behavior in favor of more exotic spin order and dynamics. In three-dimensional (3D) magnets, the pyrochlore lattice has been a particularly fruitful platform to expose new physics, in particular in rare-earth compounds.~\cite{Bramwell_2001,Gardner_2010,Fennell_2009,Ross_2011} Other three-dimensional lattice geometries, such as the diamond lattice, have been less extensively studied primarily because of the absence of obvious {geometrical frustration}.

Diamond-lattice Heisenberg antiferromagnets have attracted some recent attention, however, following the observation of a spin-liquid phase in the A-site spinel MnSc$_2$S$_4$.~\cite{Krimmel_2006,Gao_2016} This motivated detailed theoretical work that uncovered the existence of remarkable degenerate spin-spiral states when a dominant nearest-neighbor antiferromagnetic interaction competes with a small next-nearest neighbor exchange,~\cite{bergman2007order,Bernier_2008} \textit{i.e} in presence of {exchange frustration}. It was also realized that spin-orbital degeneracy may play an important role in stabilizing exotic physics as for FeSc$_2$S$_4$ \cite{Fritsch_2004,Krimmel_2005,Laurita_2015,Mittelstadt_2015,Plumb_2016,Biffin_2017} in which spin-orbital entanglement~\cite{Chen_2009a,Chen_2009b} is an active ingredient. Furthermore, as demonstrated for CoAl$_2$O$_4$ \cite{Suzuki_2007,MacDougall20092011,Zaharko_2014,MacDougall_2016}, the combination of chemical disorder with the above effects can produce unique glassy magnetic behavior of great current interest.~\cite{MacDougall20092011} 

The bipartite nature of the diamond-lattice may in fact be a favorable feature to create radically new forms of magnetism, such as the 3D topological paramagnetism recently proposed for frustrated $S\!=\!1$ diamond-lattice antiferromagnets.~\cite{Senthil_2015} In that scenario, the ground-state is an exotic superposition of fluctuating Haldane ($S\!=\!1$) chains,\cite{Haldane_1983} and can be pictured as a 3D version of the Affleck-Kennedy-Lieb-Tasaki (AKLT) construction~\cite{AKLT} used in 1D. Remarkably, NiRh$_2$O$_4$ ~\cite{chamorro2017frustrated,chen2017quantum} has already been identified as a promising candidate material to realize such topological paramagnetism, although the detailed role played by orbital degeneracy, spin-orbital entanglement, chemical disorder and exchange frustration in that material remains to be fully elucidated. 

In this paper, we focus on the antiferromagnetic A-site spinels \corho\ (cobalt rhodite) and \curho\ (copper rhodite), the latter of which is isostructural with NiRh$_2$O$_4$. Our combined experimental and theoretical work relies primarily on a neutron scattering investigation of high-quality polycrystalline samples, and establishes the canonical magnetic behavior expected for diamond-lattice Heisenberg antiferromagnets in A-site spinels. In cubic \corho\, we show that the $S\!=\!3/2$ spins are unfrustrated and display static and dynamic properties in excellent agreement with mean-field and spin-wave theory predictions. In tetragonally-distorted \curho, however, we uncover an incommensurate magnetic order for the $S\!=\!1/2$ spins and the presence of sizable quantum effects. We provide detailed modeling of these observations using mean-field and spin-wave theory up to $1/S$-order, and determine that the microscopic Hamiltonian for \curho\ involves sizable and competing magnetic exchange interactions up to the third nearest neighbor. Our results are an important reference point in the context of an accelerated search for exotic magnetic behavior on the diamond lattice.

This paper is organized as follows. Sec.~\ref{sec:methods} contains experimental details of our combined thermo-magnetic, X-ray and neutron characterization of polycrystalline samples of \corho\ and \curho. Sec.~\ref{sec:co} presents and analyzes our results on \corho, demonstrating that this coumpound is a model realization of the diamond-lattice Heisenberg antiferromagnet with $S\!=\!3/2$. Sec.~\ref{sec:cu}, discusses \curho\ for which frustrated exchange interactions lead to the development of an helical ground-state with strong zero-point reduction of the $S\!=\!1/2$ moments. In Sec.~\ref{sec:theory}, we present mean-field and spin-wave theory results for the general Hamiltonian relevant for \curho\ and discuss quantum effects in distorted diamond-lattice Heisenberg antiferromagnets that might be relevant for other materials. Sec.~\ref{sec:concl} concludes this work and additional details are provided in the Appendix.

%================  ================  ================  ================
\section{Methods}
\label{sec:methods}
%================  ================  ================  ================

%------------------------------------------
\subsection{Synthesis and determination of crystal structure}
%------------------------------------------

Black, polycrystalline samples were prepared by intimately mixing and grinding stoichiometric amounts of CoCO$_3$ (Baker Adamson, 99.9\%), CuO (Aldrich, 99.99\%), and Rh$_2$O$_3$ in an agate mortar. The Rh$_2$O$_3$ was obtained by decomposing RhCl$_3$ (Johnson Matthey, 99.9\%) at 850{\celsius} for 12 hours under air flow. The samples were then pressed as pellets and sintered at 900-950{\celsius} for 36 hours ({\curho}) and 900-1000{\celsius} for 36 hours ({\corho}) with intermediate grinding. 

Initial X-ray diffraction (XRD) characterization was performed using a Rigaku Miniflex II diffractometer using Cu K$\alpha$ radiation and a graphite monochromator. Room temperature time-of-flight neutron diffraction data were collected on POWGEN at Oak Ridge National Laboratory's (ORNL) Spallation Neutron Source (SNS) using 6-mm diameter vanadium sample cans. Rietveld analysis of the room-temperature X-ray and neutron diffraction data was carried out using the FULLPROF suite of programs.~\cite{Carvajal_1993} 

%------------------------------------------
\subsection{Thermo-magnetic measurements}
%------------------------------------------

Magnetization measurements were performed using a SQUID magnetometer in an applied magnetic field of $\mu_0 H = 0.5$~T. The temperature dependence of the magnetization $M(T)$ was measured for 2 $\leq T \leq$ 320~K on polycrystalline  samples mounted in gelatin capsules. After removing the contribution from the gelatin, the magnetic susceptibility was obtained as $\chi(T) = M(T)/H - \chi_{0}$ where $\chi_{0}=-105\times10^{-6}$~mol.emu$^{-1}$ and $\chi_{0}=-104\times10^{-6}$~mol.emu$^{-1}$ are the calculated temperature independent ionic core contributions for CoRh$_2$O$_4$ and CuRh$_2$O$_4$, respectively.~\cite{Bain_2008}

Heat capacity measurements were performed using the relaxation method on a Quantum Design Physical Properties Measurement System (PPMS) equipped with a 14~T magnet. Polycrystalline  samples were mixed with silver and pressed into pellets to increase their thermal conductivity. Contributions from the sample platform and grease, and from silver, were subtracted through separate measurements over the entire 1.6 $\leq T \leq$ 100~K temperature range of our measurements. 

%------------------------------------------
\subsection{Magnetic neutron diffraction}
%------------------------------------------

Low-temperature neutron powder diffraction measurements were performed on HB-2A at ORNL's High Flux Isotope Reactor (HFIR).~\cite{Garlea_2010} Loose polycrystalline samples (4.0~g of each of \corho\ and \curho) were enclosed in narrow 6-mm diameter cylindrical aluminum cans to minimize the effects of neutron absorption in Rh, and sealed under one atmosphere of $^4$He at room temperature. The sample cans were mounted at the bottom of a close-cycled refrigerator reaching a base temperature $T = 4$~K  and measurements were conducted with two neutron wavelengths, $\lambda = 2.41$ {\AA} from Ge(113) and $\lambda = 1.54$ {\AA} from Ge(115). 

%------------------------------------------
\subsection{Inelastic neutron scattering}
%------------------------------------------

Inelastic neutron scattering measurements were performed on the Fine-Resolution Fermi Chopper Spectrometer (SEQUOIA) at ORNL's SNS.~\cite{Granroth_2010,Stone_2014} The above samples and an empty aluminum can were mounted on a three-sample changer at the bottom of a close-cycle refrigerator reaching a base temperature of $T = 4$~K. Incident neutron energies of $E_i = 22$ meV and $E_i = 40$ meV, used in combination with a Fermi chopper frequency of 360~Hz, provided full-width at half-maximum (FWHM) elastic energy resolutions of $\delta E = 0.36$~meV and $\delta E = 0.76$~meV, respectively. Measurements were taken from base temperature to $T = 60$ K, and the contribution from the empty can has been subtracted from the inelastic neutron scattering measurements. 

%------------------------------------------
\subsection{Spin dynamics simulations}
%------------------------------------------

Unless otherwise noted, we modeled the magnetic excitations of \corho\ and \curho\ using the numerical implementation of linear spin-wave theory~\cite{Petit_2011} in the program SpinW.~\cite{Toth_2015} In our simulations, we assume a diagonal form for Heisenberg exchange interactions, \textit{i.e.} the Hamiltonian for $n$-th nearest neighbors reads $\mathcal{H}(n) = \frac{1}{2} \sum_{ij} J_n\, {\bf S}_i \cdot {\bf S}_j$ where the sum runs on all $(i,j)$ pairs of $n$-th nearest neighbor spins twice. The reported neutron scattering intensity $I(Q,E)$ for neutron energy-transfer $E\!\equiv\!\hbar \omega$ and momentum-transfer $Q\!\equiv\!|{\bf Q}|$ is proportional to the powder-averaged dynamical structure factor $S(Q,E)$ computed by SpinW, $I(Q,E) = r_0^2 |g F(Q)/2|^2\, S(Q,E)$, where $F(Q)$ is the form-factor for Co$^{2+}$ or Cu$^{2+}$ and $r_0 = 0.539 \times 10^{-12}$ cm. 

Our simulations are convoluted with a simple Gaussian lineshape to account for the $Q$ and $E$ resolution of the spectrometer, which are assumed uncoupled. The $E$-dependence of the $E$-resolution is calculated from simple geometrical considerations and calibrated with the observed elastic $E$-resolution. The  $Q$-resolution is taken to be uniform across the whole $Q$-range and estimated from the width of the observed magnetic Bragg peaks.

%================  ================  ================  ================
\section{Results on Cobalt Rhodite}
\label{sec:co}
%================  ================  ================  ================

%------------------------------------------
\subsection{Structural analysis}
%------------------------------------------

%================
\begin{figure}[t!]
	\includegraphics[width=0.99\columnwidth]{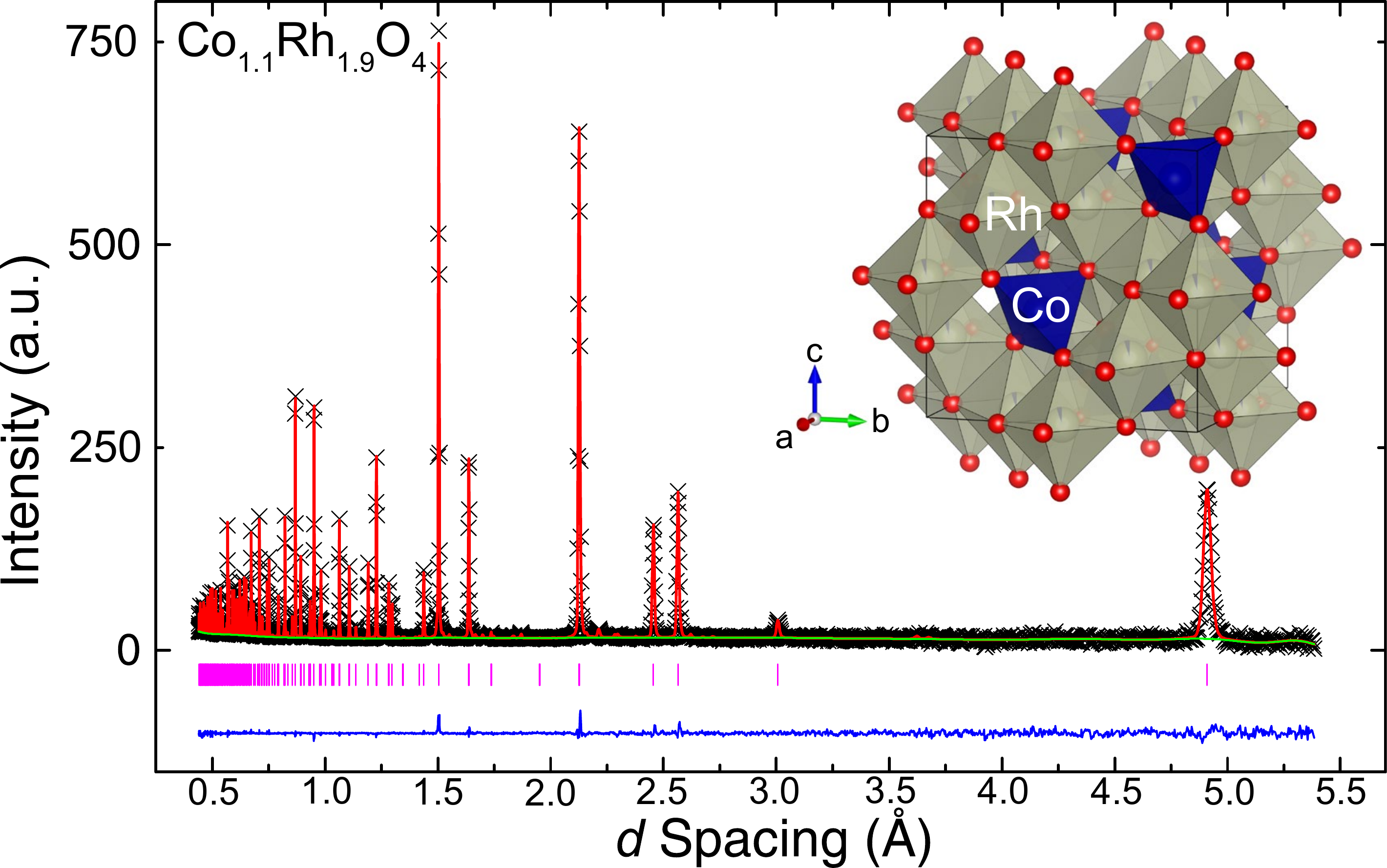} 
	\caption{(Color online) Room temperature time-of-flight neutron powder diffraction (POWGEN) results for \corho. Experimental observations are indicated by black crosses and the results of Rietveld refinements as thin lines. Vertical pink ticks indicate expected peak position and solid blue line the difference between observations and refinements. The inset depicts the crystal structures of \corho\ with O represented as red spheres, Rh octahedral with gray faces and Co tetrahedral with blue faces.}
	\label{fig:structure_co}
\end{figure}
%================

We start our experimental investigation by presenting the ideal diamond-lattice crystal structure of \corho. This material crystallizes in the cubic spinel structure [Fig.~\ref{fig:structure_co}] with space group $Fd\bar{3}m$ and room-temperature structural parameters reported in Tab.~\ref{tab:structure_co}. With respect to the general spinel structure AB$_2$O$_4$, Co$^{2+}$ occupies the tetrahedrally coordinated A-site and Rh$^{3+}$ the octahedrally coordinated B-site. This results in a perfect diamond lattice for the Co$^{2+}$ ions with four nearest-neighbor Co atoms at a distance of $3.682$~\AA. Nearest-neighbor magnetic exchange interactions are mediated by direct exchange or more likely by Co--O--Rh--O--Co superexchange paths~\cite{Blasse_1963b}. Next-nearest-neighbor exchanges, if present, involve twelve equivalent superexchange pathways with Co--Co distances of $6.013$~\AA.

%================
\begin{table}[t!]
	\renewcommand{\arraystretch}{1.5}
	\begin{center}
		%--------
		\begin{tabularx}{0.5\textwidth}{@{}l *6{>{\centering\arraybackslash}X}@{}}
		\multicolumn{7}{c}{ Co$_{1.1}$Rh$_{1.9}$O$_4$ [ $\equiv$ CoRh$_2$O$_4$  ], $T = 300$~K}\tabularnewline 
			\hline\hline
		\multicolumn{7}{c}{$Fd\bar{3}m, a\!=\!8.503(1)\mathrm{\AA}, V\!=\!614.7(1)\mathrm{\AA}^3, \chi^2\!=\!4.16, R_{\mathrm{wp}}\!=\!2.86\%$}\tabularnewline
			\hline
		{Atom} & {Site} & ${x}$ & ${y}$ & ${z}$ &{Occ.} & ${U}_{{\rm iso}}$({\AA}$^2$)\tabularnewline
			\hline
		Co & $8a$ & 0 & 0 & 0 & 1.0 & 0.0021(2)\tabularnewline
		Rh & $16d$ & 5/8 & 5/8 & 5/8 & 0.95(6) & 0.0002(1)~\tabularnewline
		Co & $16d$ & 5/8 & 5/8 & 5/8 & 0.05(6) & 0.0002(1)\tabularnewline
		O  & $32e$ & 0.2601(1) & 0.2601 & 0.2601 & 1.0 & 0.0023(1)\tabularnewline
		\hline\hline
		\end{tabularx}
		%--------
		\end{center}
		\caption{Crystallographic parameters of CoRh$_2$O$_4$ obtained by neutron powder diffraction at room temperature.} \label{tab:structure_co}
\end{table}
%================

The results of our refinement are consistent with previous reports~\cite{Bertaut_1959,Cascales_1984} with two notable differences. First, the RhO$_6$ octahedral are less distorted in our structure compared to previous reports; the shortened (respectively elongated) Co--O (respectively Rh--O) bonds lead to more chemically-reasonable bond-valence sums~\cite{Brown_1981} of 1.79 for Co and 3.05 for Rh. Second, our refinements indicate a small degree of site mixing with 5.0(6)\% of Co on the B-site and formally, a refined chemical formula of CoRh$_{1.90(1)}$Co$_{0.10(1)}$O$_4$. The Rh deficiency originates from the presence of a small Rh$_2$O$_3$ impurity phase. To maintain overall charge balance, either octahedral Co ions are $3+$, {\ie} Rh(III)$_{1.9}$Co(III)$_{0.1}$, or approximatively 5\% of the Rh ions are $4+$, {\ie} Rh(III)$_{1.8}$Rh(IV)$_{0.1}$Co(II)$_{0.1}$. Since the ionic radii for either scenario are similar it is not possible to favor one scenario over the other based on structural refinements alone. Although formally  Co$_{1.1}$Rh$_{1.9}$O$_4$, we refer to our compound as \corho\ in the rest of this manuscript.

%------------------------------------------
\subsection{Thermo-magnetic properties}
%------------------------------------------

%================
\begin{figure}[t!]
	\includegraphics[width=0.99\columnwidth]{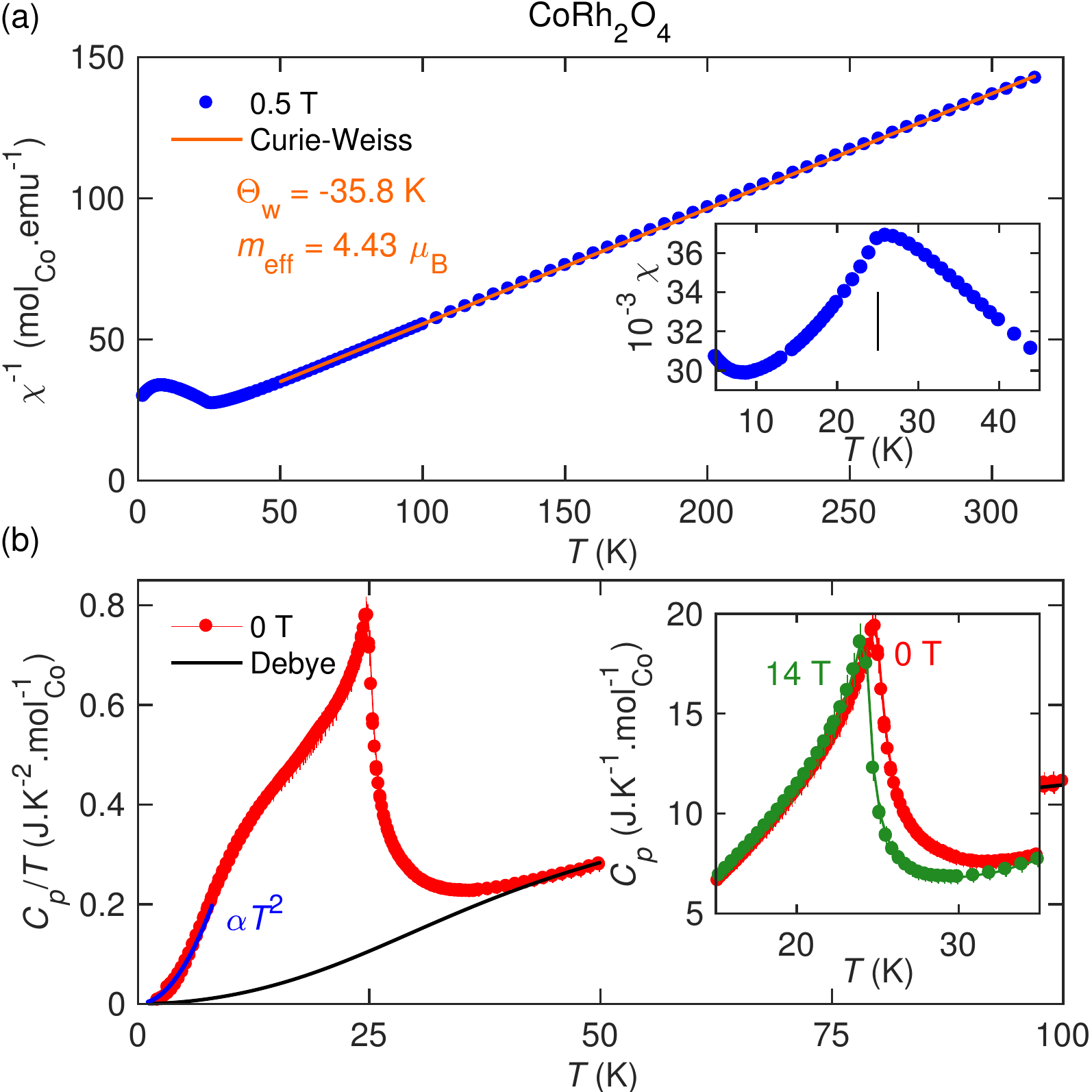} 
	\caption{(Color online) Magnetic and thermal measurements for \corho. (a) Inverse magnetic susceptibility $\chi(T)^{-1}$ in an applied magnetic field of $\mu_0H=0.5$~T (blue circles) and Curie-Weiss fit (orange line). The temperature dependence of the magnetic susceptibility $\chi(T)$ is plotted as inset with an inflection point (black vertical line) associated with the N\'eel ordering transition. (b) Temperature dependence of the total specific heat divided by temperature $C_p(T)/T$ in zero magnetic field (red circles) and Debye fit (black line) indicating the lattice contribution. The field dependence of the specific heat $C_p(T)$ around its maximum is plotted as an inset (red circles for $\mu_0H\!=\!0$~T field and green circles for $\mu_0H\!=\!14$~T).}
	\label{fig:cothermo}
\end{figure}
%================

Magnetic and thermodynamic measurements for \corho\ are presented in Fig.~\ref{fig:cothermo}. The inverse magnetic susceptibility $1/\chi(T)$ [Fig.~\ref{fig:cothermo}(a)] is linear over a broad range of temperatures $30 \leq\!T\!\leq 300$~K. A Curie-Weiss fit to the high-temperature paramagnetic regime ($T\!\geq\!50$~K) yields a negative Weiss temperature $\Theta_{\mathrm{W}} = -35.8(4)$ K and an effective moment $\mu_{\rm eff}\!=\!4.43(1)$~$\mu_{\rm B}$, consistent with previous reports.~\cite{Blasse_1963,Blasse_1963b} In the undistorted tetrahedral crystal-field environment, Co$^{2+}$ adopts the $e_g^4t_{2g}^3$ electronic configuration with one unpaired electron in each $d_{xy}$, $d_{xz}$ and $d_{yz}$ orbitals.~\cite{Bertaut_1959} For such $S=3/2$ magnetic moments, the experimental value of $\mu_{\rm eff}$ yields a gyro-magnetic ratio $g\!\approx\!2.18$ after correcting for the presence of $1.1$ Co atoms per formula unit. At low temperatures, the magnetic susceptibility $\chi(T)$ [Fig.~\ref{fig:cothermo}(a)-inset] displays a sharp absolute maximum closely followed by an inflection point at $T_{\rm N} =25.0$~K, attributed to long-range antiferromagnetic ordering.~\cite{Blasse_1963,Blasse_1963b,Fiorani_1979}

These results are fully corroborated by heat-capacity measurements. The specific heat of \corho, plotted as $C_p/T$ [Fig.~\ref{fig:cothermo}(b)], shows a sharp $\lambda$-shaped anomaly at $T_{\rm N}=25.68$~K, indicative of a second-order phase transition. The precise correspondence between specific heat and magnetic susceptibility leaves no doubt as to its magnetic nature. Most of the specific heat above $T\!\approx\!1.5 \cdot T_{\rm N}$ can be accounted for by a phonon model with two Debye temperatures, $\Theta_{\rm D}={253(3)}$~K and 742(9)~K. Integrating the magnetic part of $C_p/T$ from 1.7~K to 50~K yields an entropy change $\Delta S\!\approx\!11.7$~J.K$^{-1}$.mol$^{-1}$, consistent with $R\ln 4\!=\!11.52$~J.K$^{-1}$.mol$^{-1}$ expected for $S\!=\!3/2$ degrees of freedom. Below $T_{\rm N}$, the magnetic contribution to the specific heat dominates and a broad feature is observed around $T^*\!\approx\!12$~K, which we attribute to magnon-magnon interactions. Below $T^*$, the specific heat follows a $C_p\!=\!\alpha T^3$ behavior, as expected for gapless antiferromagnetic magnons. Given the relatively large energy scale set by $\Theta_{\mathrm{W}}\!\approx\!35$~K, a large applied magnetic field of $\mu_0 H\!=\!14$~T has almost no influence on the transition temperature. We observe a shift downward by a mere $0.74$~K [Fig.~\ref{fig:cothermo}(b)-inset]. Overall, our measurements yield a frustration ratio~\cite{Ramirez_1994} $f\!=\!|\Theta_{\rm W}|/T_{\rm N}\!=\!1.4$ and suggest that \corho\, behaves as a canonical non-frustrated three-dimensional antiferromagnet with an average exchange interaction between nearest-neighbor magnetic moments ($z\!=\!4$) of $J_{\rm av}\!=\!3 k_{\rm B}\Theta_{\rm W}/z S (S+1)\!=\!0.62$~meV.

%------------------------------------------
\subsection{Magnetic structure}
%------------------------------------------

%================
\begin{figure}[t!]
	\includegraphics[width=0.90\columnwidth]{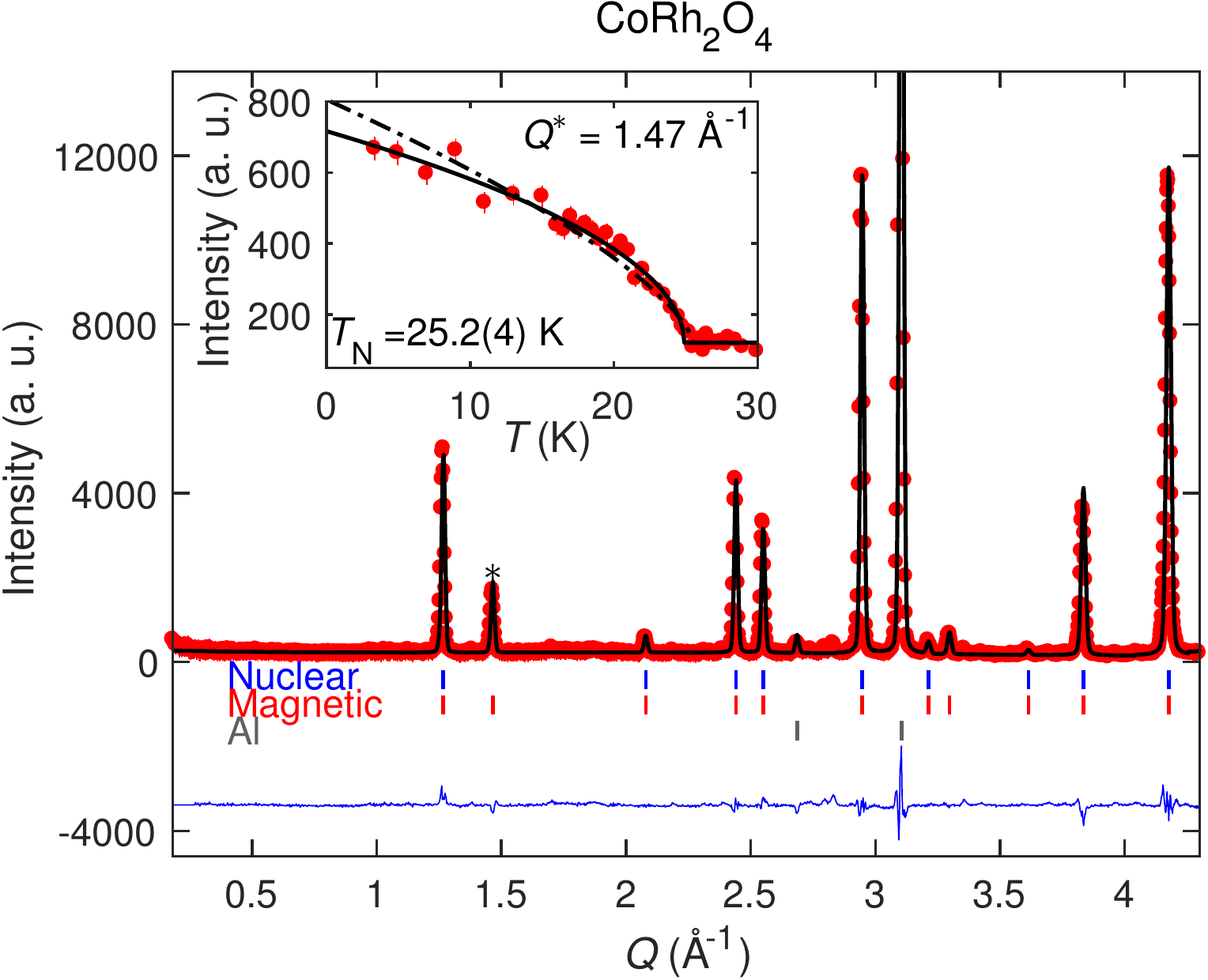} 
	\caption{(Color online) Neutron powder diffraction (HB-2A) pattern of \corho\ measured at $T\!=\!4$ K with a neutron wave-length of $\lambda= 2.41$\,\AA\  (red circles), and corresponding Rietveld refinements (black line) of the nuclear (blue ticks), magnetic (red ticks) and aluminum background (gray ticks) contributions.  The blue line shows the difference between data and the best Rietveld fit for which peak shapes were modeled using a pseudo-Voigt function. The inset shows the temperature dependence of the most intense magnetic peak (highlighted with an asterisk). The black solid (respectively dot-dashed) curve is an order parameter fit with a fixed mean-field (3+1D) Ising (resp. Heisenberg) critical exponent $\beta = 0.25$ (resp. $\beta = 0.34$) in order to estimate the value of $T_{\rm N}$.}
	\label{fig:corietveld}
\end{figure}
%================

Neutron powder diffraction allows one to determine the magnetic structure of \corho\ below the antiferromagnetic ordering transition at $T_{\rm N} \approx 25$~K [Fig.~\ref{fig:corietveld}]. Upon cooling our sample from 40\,K to 4\,K, we observe a sizable change of intensity for some of the nuclear Bragg peaks, coinciding with the development of new Bragg peaks at nuclear positions forbidden by the space-group symmetry, for instance $\{h,k,\ell\}=\{2,0,0\}$ ($Q^*\!=\!1.47$~\AA$^{-1}$) and $\{2,4,0\}$ ($Q\!=\!3.30$~\AA$^{-1}$). The integrated intensity of the $\{2,0,0\}$ peak [Fig.~\ref{fig:corietveld}-inset] follows an order-parameter behavior with a sharp onset at $T_{\mathrm{N}} = 25.2(4)$\,K, in close correspondence with the thermodynamic anomalies. We thus associate the change in Bragg scattering with the development of long-range magnetic ordering.

%================
\begin{figure}[t!]
	\includegraphics[width=0.99\columnwidth]{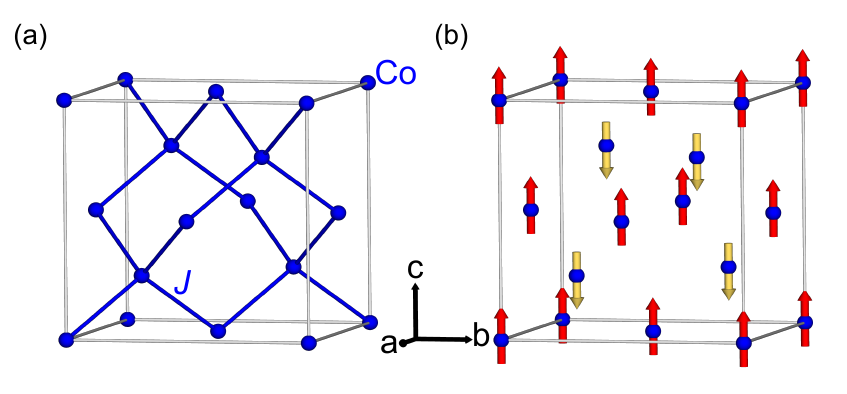} 
	\caption{(Color online) Conventional body-centered unit cell for Co atoms (blue spheres) in \corho\ showing (a) the diamond-lattice connectivity of the nearest neighbor bonds (blue lines) and (b) the two sublattice $\mathbf{k}_{\rm m}=(0,0,0)$ antiferromagnetic magnetic structure (red and yellow arrows) obtained from our Rietveld refinement.}
	\label{fig:comag}
\end{figure}
%================

All the observed magnetic Bragg peaks can be indexed by the magnetic propagation vector $\mathbf{k}_{\rm m}=(0,0,0)$ with respect to the conventional unit cell. To determine the magnetic structure, we first investigate possible symmetry-allowed magnetic structures using the program Isodistort.~\cite{Campbell_2006} For CoRh$_2$O$_4$, there are two irreducible representations (irreps), labeled $\Gamma_{4+}$ and $\Gamma_{5-}$ in the notation of Miller and Love.~\cite{Miller_1967} These correspond to simple ferromagnetic and antiferromagnetic ordered pattern on the diamond lattice [Fig.~\ref{fig:comag}(a)], respectively. As anticipated from the negative Curie-Weiss constant, only $\Gamma_{5-}$ correctly accounts for the observed magnetic intensity. The resulting spin structure (magnetic space group $I4_{1}{^\prime}/a{^\prime}m{^\prime}d$) is shown in Fig.~\ref{fig:comag}(b). Our Rietveld refinement [Fig.~\ref{fig:corietveld}] is in excellent agreement with the data ($R_\mathrm{wp}\!=\!8.11\%, R_\mathrm{mag}\!=\!7.62\%$) and yields an ordered magnetic moment $\mu_{\mathrm{ord}} =3.11(5)\,\mu_{\mathrm{B}}$, close to the value of $gS = 3.27\,\mu_{\mathrm{B}}$ expected for a $S=3/2$ ion with $g=2.18$. Neutron powder diffraction thus demonstrates that \corho\ orders in a simple two-sublattice antiferromagnetic structure at $T_{\rm N}\!\approx\!25$~K and places an upper bound of 5\% on any reduction of the ordered moment due to quantum fluctuations at $T\!=\!4$~K.

%------------------------------------------
\subsection{Magnetic excitations}
%------------------------------------------

\begin{figure}[t!]
\renewcommand{\arraystretch}{1.3}
	%\begin{tabular}{cc}
	%\hline\hline
	%{\corho} \hspace{2cm} & $J_1 = 0.63$ meV \tabularnewline \hline \hline
	%\end{tabular}
	%\\[5px]
	\includegraphics[width=0.99\columnwidth]{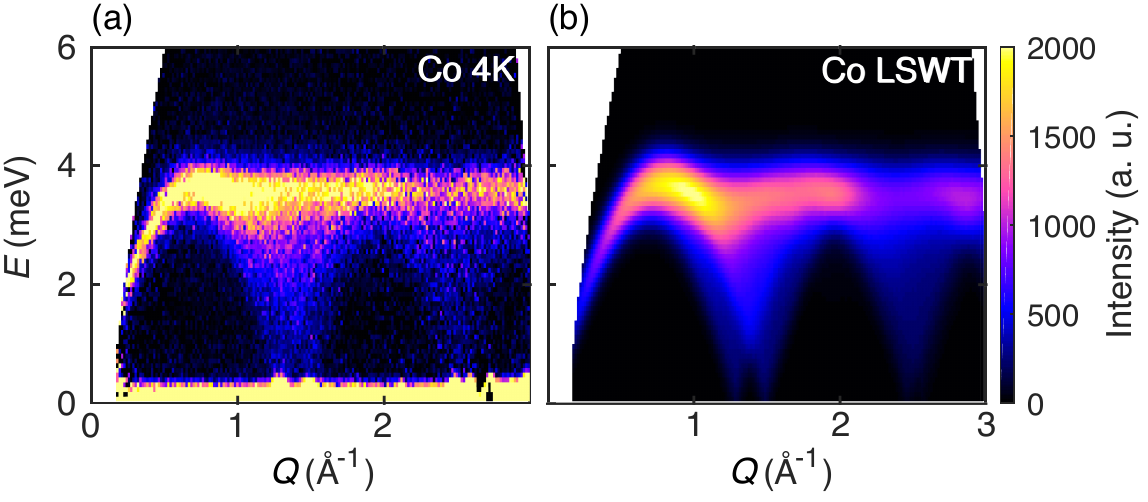}
	\caption{(Color online) Magnetic excitations of \corho. (a) Momentum and energy  dependence of the powder inelastic neutron scattering intensity $I(Q,E)$ at $T=4$~K. (b) Linear spin-wave theory simulations of $I(Q,E)$, for the magnetic structure of Fig.~\ref{fig:comag}(b), stabilized by a nearest-neighbor exchange interaction $J_1=0.63$~meV.} 
	\label{fig:coinelastic}
\end{figure}

Inelastic neutron scattering measurements on \corho\ [Fig.~\ref{fig:coinelastic}(a)] reveal a simple magnetic excitation spectrum we associate with non-interacting magnons, \textit{i.e.} spin fluctuations transverse to the ordered spin patterns of Fig.~\ref{fig:comag}(b). The magnetic spectrum appears gapless within the resolution of our experiments, with characteristic acoustic spin-wave branches emerging from the strong magnetic Bragg peak positions. The bandwidth of the magnetic signal $W\approx3.8$~meV = 44~K matches well with the value of the Weiss constant $\Theta_{\rm W}\!=\!35.5$~K and corresponds to the energy of magnons at the Brillouin zone boundary. We obtain an excellent correspondence between the data and the calculated scattering intensity [Fig.~\ref{fig:coinelastic}(b)] with a single nearest-neighbor exchange parameter $J_1\!=\!0.63$~meV [Fig.~\ref{fig:comag}(b)]. This matches very well with the average exchange value extracted from the magnetic susceptibility $J_{\rm av}\!=0.62$~meV, indicating that further neighbor exchanges and $T\!=\!0$ magnon energy renormalization effects can be neglected in \corho.

%================
%================

%================
\begin{figure*}
	\begin{minipage}{0.99\textwidth}
		\includegraphics[width=0.99\textwidth]{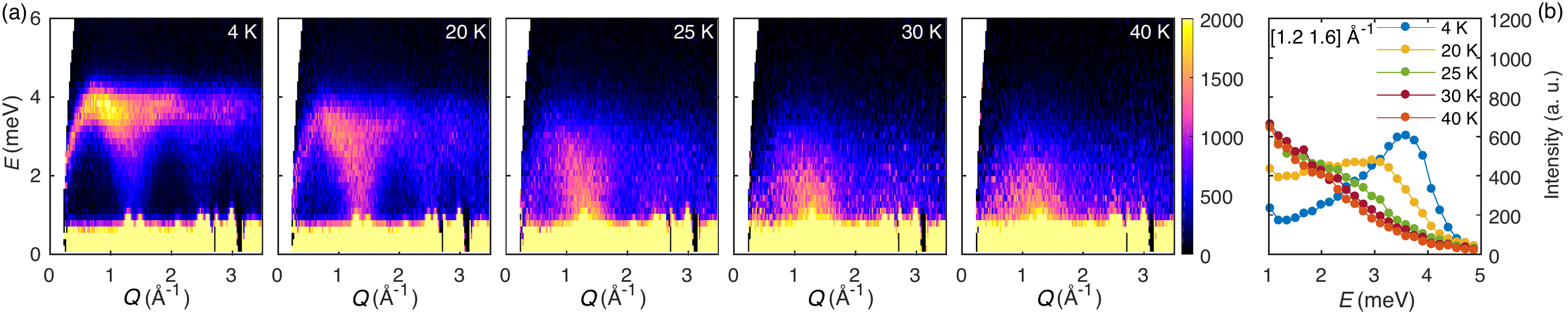}	
	\end{minipage}
		\caption{(Color online) Temperature dependence of the magnetic excitations of \corho. (a) Evolution of the $Q$--$E$ scattering intensity upon crossing $T_{\rm N}\approx25$~K. (b) Constant-$Q$ cut around the ordering wave-vector position $Q^*=1.4\pm0.2~$\AA$^{-1}$.}
	\label{fig:cotdep}
\end{figure*}
%================

The temperature dependence of the magnetic excitations [Fig.~\ref{fig:cotdep}(a)] reveals a very rapid collapse of the magnetic excitations as $T_{\rm N}$ is crossed. Unlike low-dimensional quasi-1D and quasi-2D magnets for which the overall bandwidth and shape of the magnetic excitations persists at and above $T_{\rm N}$,~\cite{Lake_2005,Ronnow_1999} the excitations of \corho\ resemble that of a paramagnet already for $T\!\gtrapprox\!T_{\rm N}$. The top of the magnon band is considerably renormalized and broadened at $T=T_{\rm N}$ , a temperature above which the excitations loose coherence and the inelastic signal becomes purely relaxational [Fig.~\ref{fig:cotdep}(b)]. While the detailed analysis of the temperature dependence of these excitations is beyond the scope of this work, the simplicity of the $T \ll T_{\rm N}$ spectrum and the presence of an unique energy scale $J_1\!=\!0.63$~meV makes \corho\ a model 3D antiferromagnetic material.

%================  ================  ================  ================
\section{Results on Copper Rhodite}
\label{sec:cu}
%================  ================  ================  ================

%------------------------------------------
\subsection{Structural analysis}
%------------------------------------------

%================
\begin{figure}
	\includegraphics[width=0.99\columnwidth]{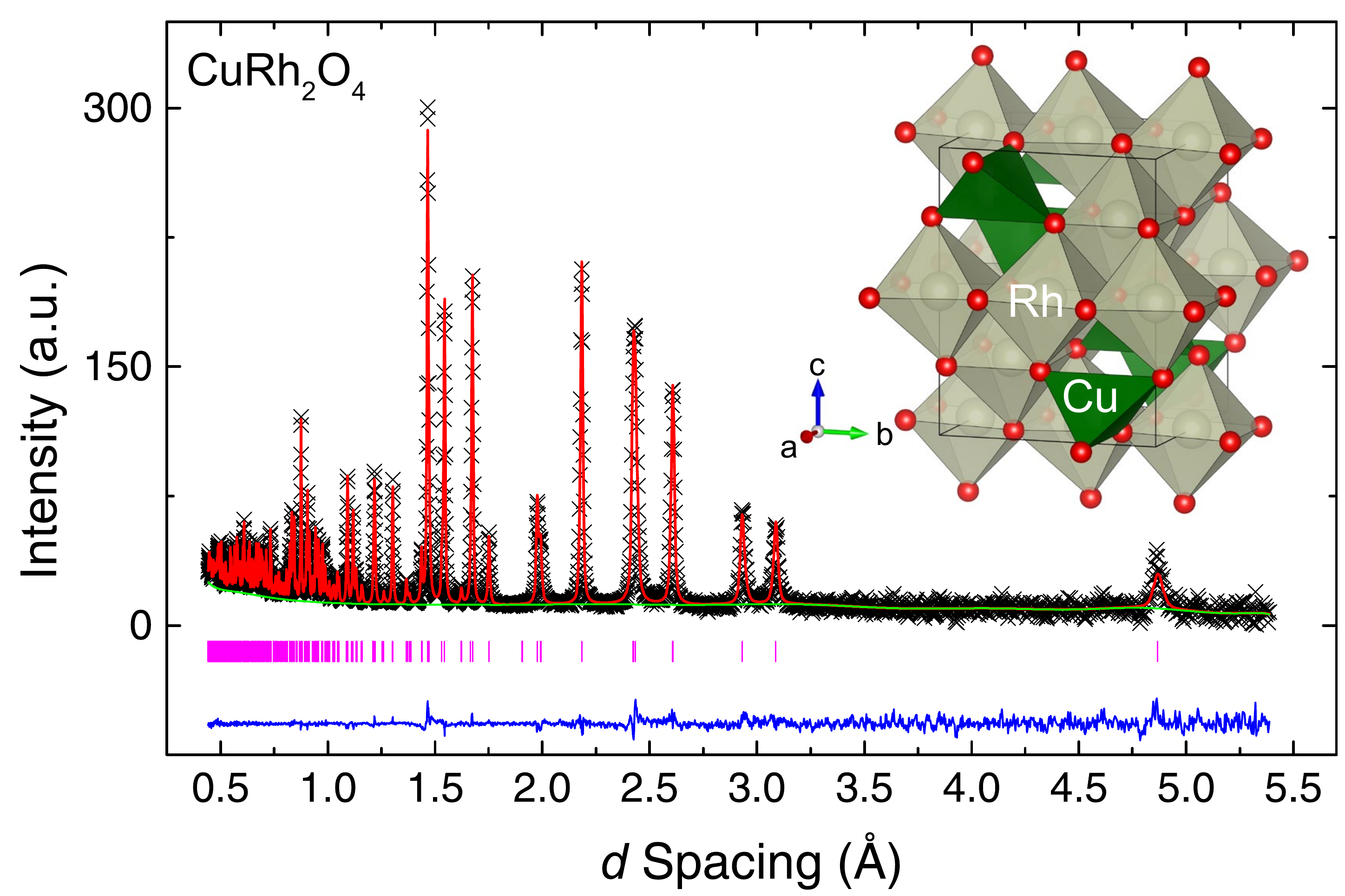} 
	\caption{(Color online) Room temperature time-of-flight neutron powder diffraction (POWGEN) results for \curho. Experimental observations are indicated by black crosses and the results of Rietveld refinements as thin lines. Vertical pink thicks indicate expected peak position and solid blue line the difference between observations and refinements. The inset depicts the crystal structure of \curho\ with O represented as red spheres, Rh octahedral with gray faces and Cu tetrahedral with green faces.\\}
	\label{fig:structure_cu}
\end{figure}
%================

{\curho} crystallizes in a lower-symmetry crystal structure than \corho\ due to a Jahn-Teller distortion around $T_{\rm JT}\approx850$~K~\cite{Bertaut_1959,Blasse_1963b} lifting the degeneracy of the $e_g^4t_{2g}^5$ electronic configuration of Cu$^{2+}$. The necessary destabilization of the magnetic $d_{xy}$ orbital below $T_{\rm JT}$ leads to a compression of the oxygen tetrahedral with respect to the cubic cell.~\cite{Bertaut_1959} Indeed the structure of \curho\ has been described by both X-ray \cite{Khanolkar_1961} and neutron diffraction \cite{Kennedy_1999} as a tetragonally distorted spinel with space group $I4_1/amd$~\cite{Kennedy_1999} or $I\bar{4}2d$.\cite{Khanolkar_1961}

%================
\begin{table}[b!]
	\renewcommand{\arraystretch}{1.3}
	\begin{center}
		%--------
		\begin{tabularx}{0.5\textwidth}{@{}l *6{>{\centering\arraybackslash}X}@{}}
		\multicolumn{7}{c}{\curho, $T=300$~K}\tabularnewline 
		\hline\hline
		\multicolumn{7}{c}{ $I{4}_{1}/amd, a\!=\!6.177(1)\mathrm{\AA}, c\!=\!7.902(1)\mathrm{\AA}, V\!=\!301.5(1)\mathrm{\AA}^3$} \tabularnewline
		\multicolumn{7}{c}{$\chi^2\!=\!3.86, R_{\mathrm{wp}}\!=\!2.66\%$}\tabularnewline
		\hline
		{Atom} & {Site} & ${x}$ & ${y}$ & ${z}$ &{Occ.} & ${U}_{{\rm iso}}$({\AA}$^2$)\tabularnewline
		\hline
		Cu & $8e$ & 0 & 3/4 & 0.1368(3) & 1.0 & 0.0017(2)\tabularnewline
		Rh & $8d$ & 0 & 0 & 1/2 & 1.0 & 0.0005(1)\tabularnewline
		O & $16h$ & 0 & 0.0334(1) & 0.2430(1) & 1.0 & -\tabularnewline[5px]	
		\hline	
        \multicolumn{7}{c}{{Anisotropic Atomic Displacement Parameters ({\AA}$^2$)}}\tabularnewline
 		\hline
 		{Atom} & ${U}_{{11}}$ & ${U}_{{22}}$ & ${U}_{{33}}$ & ${U}_{{12}}$ & ${U}_{{13}}$ & ${U}_{{23}}$\tabularnewline
 		\hline
		O & 0.0021 & 0.0010 & 0.0022 & 0.0 & 0.0 & $0.0002$\tabularnewline		
		\hline\hline
		\end{tabularx}
		%--------
	\end{center}
		\caption{Crystallographic parameters of CuRh$_2$O$_4$ obtained by neutron powder diffraction at room temperature.} \label{tab:structure_cu}
\end{table}%
%================

Our room temperature neutron diffraction results for {\curho} are shown in Fig.~\ref{fig:structure_cu}. The results of our Rietveld refinement, reported in Table~\ref{tab:structure_cu}, yield $I4_1/amd$ as the appropriate room-temperature space group, consistent with the most recent studies.~\cite{Kennedy_1999,Dollase_1997} Unlike \corho, we find no evidence for site mixing with bond valence sums of 3.05 for Rh, 1.97 for O and 1.79 for Cu. A close look at the crystal structure indicates that Rh octahedral are distorted with four distinct O---Rh---O bond angles of 98.23(4)$^{\circ}$, 81.77(4)$^{\circ}$, 92.83(5)$^{\circ}$ and 87.17(5)$^{\circ}$. In turn, the Cu tetrahedral are flattened with two distinct O---Cu---O bond angles of 128.8(2)$^{\circ}$ and 102.6(1)$^{\circ}$. For comparison, there are only two O-Rh-O angles of 85.10(3)$^{\circ}$ and 94.90(3)$^{\circ}$ and a single O-Cu-O angle of 109.47(3)$^{\circ}$ in \corho. Our refined crystal structure also indicates Cu is displaced off the ideal $4a$ site in a disordered manner. Instead, the copper position splits between two $8e$ positions that are randomly occupied along the $c$ axis. Overall the tetragonal distortion leads to four nearest-neighbor Cu---Cu distances within $0.2\%$ of each other such that nearest-neighbor Cu$^{2+}$ ions in \curho\  effectively remain organized on a diamond lattice at an average distance of 3.61(5)~\AA. When compared to the cubic structure of \corho, however, next-nearest-neighbor Cu---Cu distances are strongly split into four short and eight long links. We will see below this has profound consequences for the magnetic properties of \curho.

%------------------------------------------
\subsection{Thermo-magnetic properties}
%------------------------------------------

\begin{figure}
	\includegraphics[width=0.99\columnwidth]{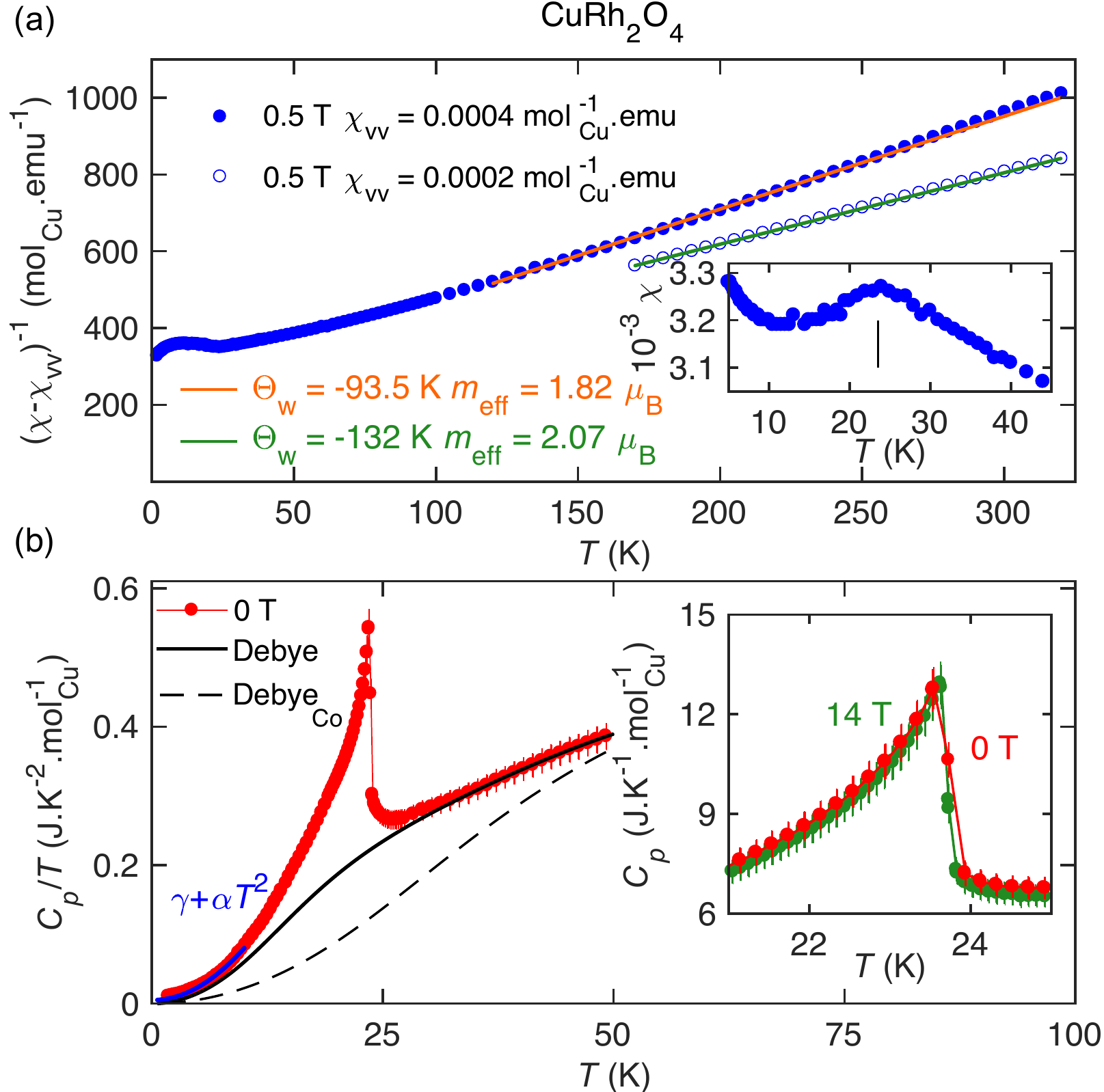} 
	\caption{(Color online) Magnetic and thermal measurements for \curho. (a) Inverse magnetic susceptibility $(\chi(T)-\chi_{\rm VV})^{-1}$ in an applied magnetic field of 0.5 T with two distinct temperature-independent Van-Vleck contributions $\chi_{\rm VV}$ subtracted (solid and empty blue circles) and corresponding Curie-Weiss fits (orange and green lines, respectively). Temperature dependence of the total specific heat divided temperature $C_p(T)/T$ in zero magnetic field (red circles) and fits to the lattice contribution of \curho\ (black line) and \corho\ (dashed line). The field dependence of the specific heat $C_p(T)$ around its maximum is plotted as inset (red circles for $\mu_0H=0$~T field and green circles for $\mu_0H=14$~T.}
	\label{fig:cuthermo}
\end{figure}

Magnetic and thermodynamic measurements for \curho\ are presented in Fig.~\ref{fig:cuthermo}. Unlike \corho\, the inverse magnetic susceptibility $1/\chi(T)$ [Fig.~\ref{fig:cuthermo}(a)] only becomes linear at high temperature after subtraction of a positive Van-Vleck contribution $\chi_{\rm VV}$, associated with paramagnetic Rh$^{3+}$.~\cite{Endoh_1999} Linearity of $[\chi(T)-\chi_{\rm VV}]^{-1}$ for $T\geq170$~K is obtained using $\chi_{\rm VV} = +200 \times 10^{-6}$~emu.mol$^{-1}$ from which a Curie-Weiss fit yields $\Theta_{\mathrm{W}} = -132.2(7)$~K and $\mu_{\rm eff}\!=\!2.073(2)$~$\mu_{\rm B}$. The obtained effective moment is somewhat too large for Cu$^{2+}$. Using an empirical $\chi_{\rm VV} = +400 \times 10^{-6}$~emu.mol$^{-1}$, we obtain a good Curie-Weiss fit above $T\geq120$~K with values of $\Theta_{\mathrm{W}} = -93(1)$~K and $\mu_{\rm eff}\!=\!1.819(5)$~$\mu_{\rm B}$, compatible with a previous report~\cite{Endoh_1999} and corresponding to a realistic gyro-magnetic ratio $g\!\approx\!2.1$ for the Cu$^{2+}$ ions. At low temperatures, the magnetic susceptibility  $\chi(T)$ [Fig.~\ref{fig:cuthermo}-inset] displays a local maximum with an inflection point at $T_{\rm N}\!=\!23.5$~K, indicating antiferromagnetic ordering.~\cite{Endoh_1999}

%================
%================

The specific heat of \curho\ [Fig.~\ref{fig:cuthermo}(b)] displays a sharp $\lambda$-shaped peak at $T_{\rm N}\!=\!23.54$~K in perfect correspondence with the susceptibility result. This peak shifts by less than $0.1$~K when a magnetic field of $\mu_0{\it H}\!=\!14$~T is applied [Fig.~\ref{fig:cuthermo}(b)-inset]. A phonon model with two Debye temperatures, $\Theta_{\rm D}=116(12)$~K and $350(14)$~K, accounts for most of the specific heat for $T\geq1.5 \cdot T_{\rm N}$ but overestimates the phonon contribution as the entropy change from 1.7~K to 50~K, $\Delta S\!\approx\! 3.2$~J.K$^{-1}$.mol$^{-1}$, falls short of $R\ln 2\!=\!5.76$~J.K$^{-1}$.mol$^{-1}$ expected for $S\!=\!1/2$ degrees of freedom. Using the Debye model from \corho\, the magnetic entropy reaches $\Delta S\!\approx\!6.5$~J.K$^{-1}$.mol$^{-1}$ at 50~K, the large value of which suggests possible magneto-elastic effects. Below $T\leq20$~K, the specific heat is well described by $C_p\!=\!\gamma T +\alpha T^3$, where the small $\gamma\!=\!5\times10^{-3}$ J.K$^{-2}$.mol$^{-1}$ term may indicate weak glassiness in the low energy spectrum of otherwise gapless antiferromagnetic magnons. The large $\Theta_{\rm W}$ compared to $T_{\rm N}$ suggests a moderate degree of frustration in \curho, with $4.0\!\leq\!f\!\leq\!5.5$. In the following, we investigate the nature and consequences of competing (frustrated) exchange interactions in \curho.

%------------------------------------------
\subsection{Magnetic structure}
%------------------------------------------

\begin{figure}
	\includegraphics[width=0.88\columnwidth]{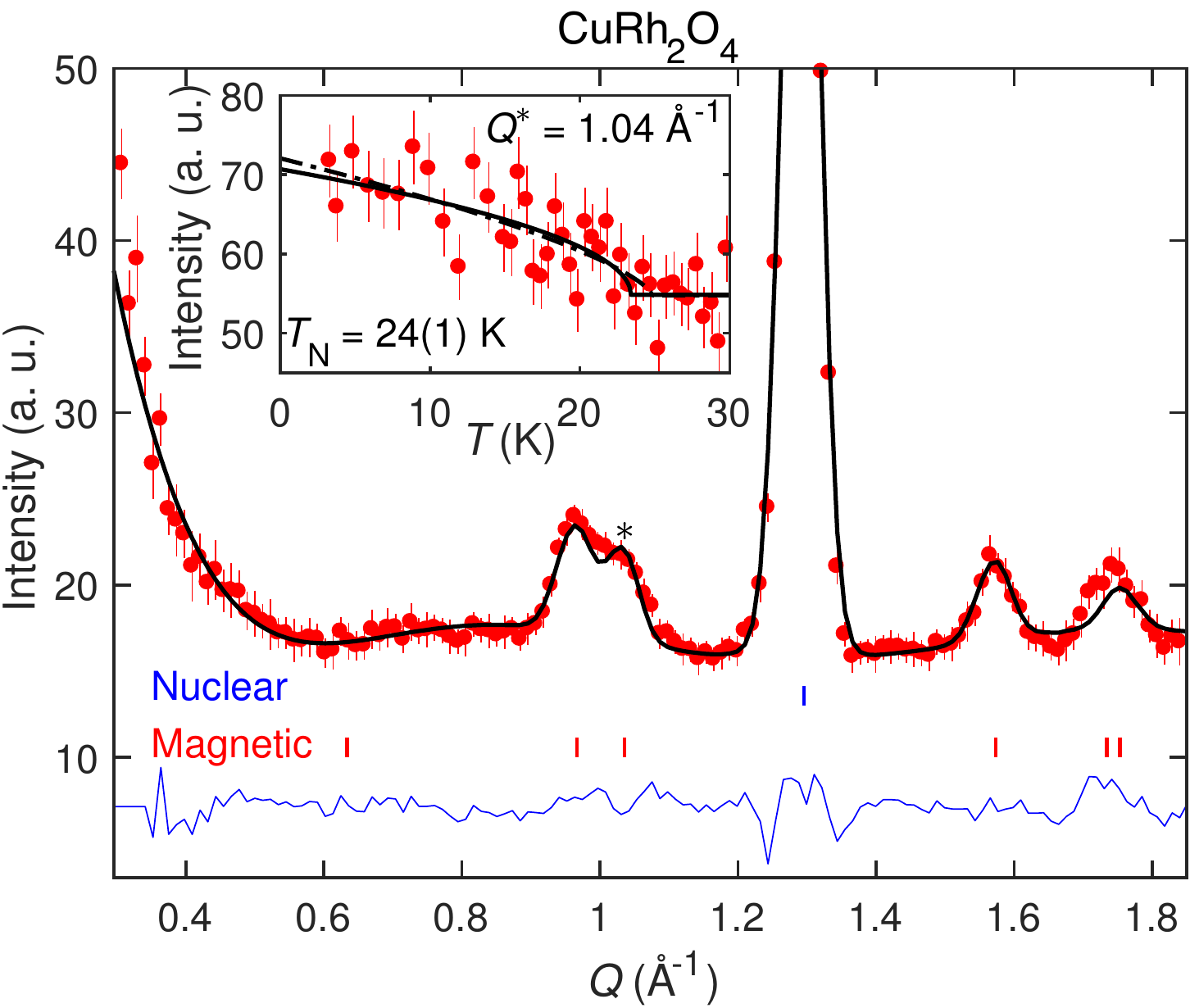} 
	\caption{(Color online) Rietveld refinement of our {\curho} elastic scattering results at $T\!=\!4$ K, extracted from an $E$-integrated elastic cut through our inelastic data (SEQUOIA). Nuclear (blue ticks) and magnetic (red ticks) phases are included. The inset shows the temperature dependence of one of the most intensive magnetic peak (highlighted with asterisk) obtained by neutron diffraction (HB-2A). The black solid (respectively dot-dashed) curve is an order parameter fit with a fixed mean-field (3+1D) Ising (resp. Heisenberg) critical exponent $\beta = 0.25$ (resp. $\beta = 0.344$) in order to estimate the value of $T_{\rm N}$.}
	\label{fig:curietveld}
\end{figure}

More direct evidence for the presence of frustration in \curho\ comes from low-temperature neutron diffraction. After cooling our sample of CuRh$_2$O$_4$ from 25\,K to 4\,K [Fig.~\ref{fig:curietveld}], we observed new Bragg peaks at small wave-vectors ($Q\lesssim 2$\,\AA$^{-1}$). Given the known thermodynamic anomalies, we identify these peaks with the development of long-range magnetic order. As anticipated for a $S\!=\!1/2$ system, these magnetic Bragg peaks are very weak. In fact, we observed only a single magnetic peak above background (at $Q=1.04$\,\AA$^{-1}$) in our diffraction data taken with $\lambda= 2.41$\,\AA\ and optimized for high resolution. The temperature dependence of the integrated intensity of this peak [Fig.~\ref{fig:curietveld}-inset] yields $T_{\mathrm{N}} = 24(1)$\,K. However, we were able to observe several magnetic Bragg peaks with good statistics by integrating our inelastic scattering data over the elastic energy resolution [Fig.~\ref{fig:curietveld}], which we will henceforth refer to as ``elastic scattering''.

%================
%================

\begin{figure}
	\includegraphics[width=0.99\columnwidth]{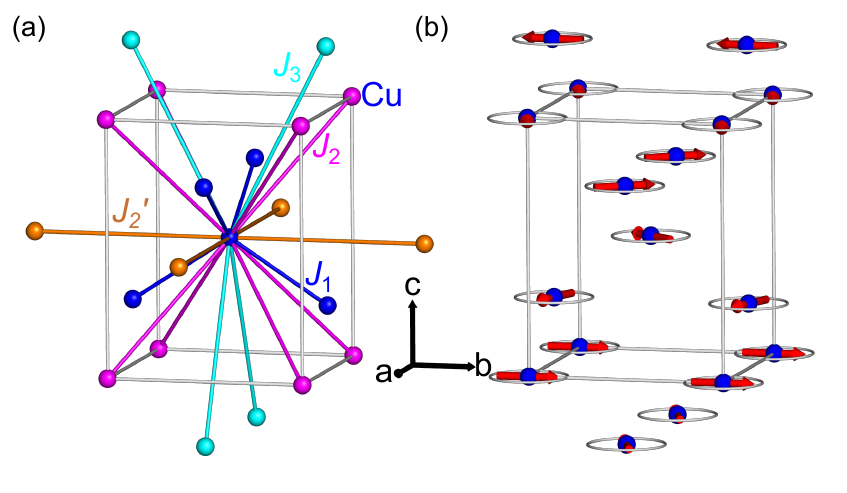} 
	\caption{Conventional body-centered unit cell for Cu atoms (colored spheres) in \curho\ showing (a) the connectivity of the nearest neighbor (blue lines), second neighbor (pink and orange lines), and third neighbor bonds (cyan lines). The two distinct kinds of second neighbor interactions are degenerate in the cubic case. (b) Incommensurate magnetic structure of \curho\ with $\mathbf{k}_{\rm m}=(0,0,0.79)$ and $\cos\varphi=-0.32$.}
	\label{fig:cumag}
\end{figure}

The magnetic Bragg peaks are indexed by an incommensurate magnetic propagation vector $\mathbf{k}_{\rm m}=(0,0,k_z)$ with respect to the conventional unit cell, where $k_z\!\approx\!0.79$. For the space-group of \curho\ and $\mathbf{k}_{\rm m}$, there are three irreps, of which two are one-dimensional, $\Lambda_3$ and $\Lambda_4$, and one is two-dimensional, $\Lambda_5$.~\cite{Campbell_2006} However, the one-dimensional irreps can be discounted, because they correspond to amplitude-modulated spin-density waves with the ordered magnetic moment parallel to the $\mathbf{c}$ axis of the tetragonal unit cell, which would lead to the $\{0,0,2-k_z\}$ Bragg peak ($Q\!=\!0.962$~\AA$^{-1}$) being absent, in conflict with experimental observations. The $\Lambda_5$ irrep corresponds to the ordered spin component lying in the $ab$ plane and it contains two candidate magnetic structures for which all spins possess ordered magnetic moments of equal magnitude. Both structures are circular helices ($\mathbf{k}_{\rm m}$ perpendicular to the spins' plane of rotation), with the angle $\varphi$ between adjacent spins along ${\bf c}$ given by $\cos\varphi=\pm0.32$. Calculating the powder-diffraction patterns reveals that only the structure with $\cos\varphi\!=\!-0.32$ shows good agreement with experimental data. We therefore identify the magnetic structure of CuRh$_2$O$_4$ as a circular helix with $\cos\varphi\!=\!-0.32$. This structure (magnetic space group $I4_{1}221^{\prime}$), which probably originates from competing exchange interactions [Fig.~\ref{fig:cumag}(a)], is shown in Fig.~\ref{fig:cumag}(b).

We performed Rietveld refinements against our neutron data to obtain accurate values for $k_z$ and the ordered magnetic moment length $\mu_{\mathrm{ord}}$. Because the elastic data have high statistics but relatively low resolution, while the opposite is true of the diffraction data, we fit to several datasets simultaneously; namely, the $4-25$\,K elastic data (magnetic phase), the $4$\,K elastic data (magnetic and nuclear phases), the $4-25$\,K diffraction data (magnetic phase), and the $4$\,K diffraction data (nuclear phase). The magnetic phase was excluded from the fit to the $4$\,K diffraction data because of additional weak peaks from the sample environment, which may bias the magnetic refinement. The fit to the 4\,K elastic data [Fig.~\ref{fig:curietveld}] represents good agreement with the data ($R_\mathrm{wp} =6.53\%; R_\mathrm{mag} = 18.8\%$). The refined parameter values are $k_z\!=\!0.790(4)$ and $\mu_{\mathrm{ord}}\!=\!0.56(6)\,\mu_{\mathrm{B}}$. The value of $\mu_{\mathrm{ord}}$ is significantly reduced from its maximum expected value of $1.05\,\mu_{\mathrm{B}}$, which indicates strong quantum fluctuations, an effect we consider in detail below.

%================
%================

%------------------------------------------
\subsection{Magnetic excitations}
%------------------------------------------

%================
\begin{figure}
\renewcommand{\arraystretch}{1.5}
	\begin{tabular}{cll}
	\hline\hline
	\hspace{0.7cm} \curho \hspace{0.7cm} & $J_1 = 10.1$ meV  & $J_2 = 0.32~J_1$ \\
	 	   & $J_3 = 0.098~J_1 $  & $J^\prime_2 = 0.18~J_1 $ \\
	 	   \hline\hline
	\end{tabular}\\[5px]
	\includegraphics[width=0.99\columnwidth]{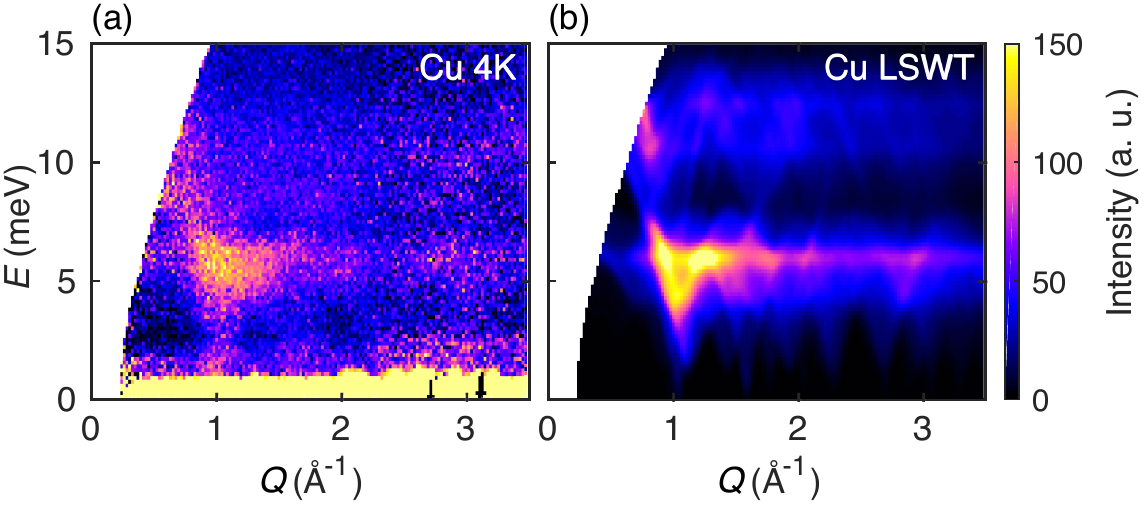}
	\caption{(Color online) Magnetic excitations of \curho. (a) Momentum and energy  dependence of the powder inelastic neutron scattering intensity $I(Q,E)$ at $T=4$~K. (b) Linear spin-wave theory simulations of $I(Q,E)$, for the magnetic structure of Fig.~\ref{fig:cumag}(b), stabilized by the magnetic exchange interactions listed above the plot and defined in Fig.~\ref{fig:cumag}(a).}
	\label{fig:cuinelastic}
\end{figure}
%================

%================
\begin{figure*}
	\begin{minipage}{0.99\textwidth}
		\includegraphics[width=0.99\textwidth]{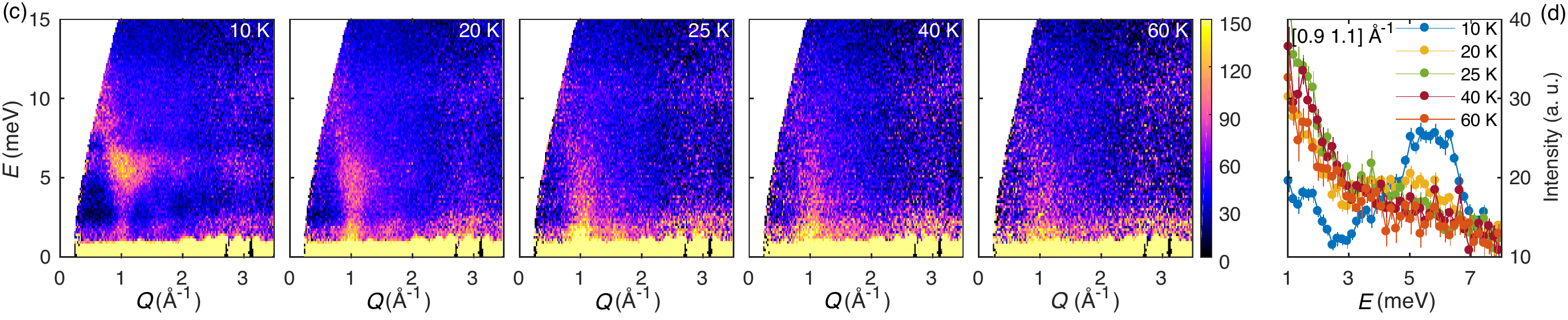}	
	\end{minipage}
		\caption{(Color online) Temperature dependence of the magnetic excitations of \curho. (a) Evolution of the $Q$--$E$ scattering intensity upon crossing $T_{\rm N}\approx23$~K. (b) Constant-$Q$ cut around the ordering wave-vector position $Q=1.0\pm0.1~$\AA$^{-1}$.}
	\label{fig:cutdep}
\end{figure*}
%================

To explain the origin of this incommensurate magnetic structure, we resort to inelastic neutron scattering to determine the values of possible magnetic exchange interactions for the distorted structure of \curho\ [Fig.~\ref{fig:cumag}(a)]. The magnetic spectrum of \curho\ appears gapless within the resolution of our experiments but unlike \corho\ we observe a complex landscape of high-energy excitations, with peaks in the density of magnetic scattering at $W_1\!=\!6.2\!$~meV and $W_2\!\approx\!11.5\!$~meV, several times greater than the excitation bandwidth of \corho. Given that the nearest-neighbor magnetic ion distances  are very similar for the two compounds (3.68~\AA\ and 3.61~\AA), this suggests super-exchange interactions very sensitive to the details of the crystal structure. Furthermore, the presence of two apparent energy scales in \curho\ implies that several exchange interactions exist, and potentially compete, to stabilize the incommensurate magnetic structure.

To model the excitations of \curho, we consider a Heisenberg model with up to third-nearest neighbor interactions; see Fig.~\ref{fig:cumag}(a). The nearest-neighbor interaction $J_1$ defines a diamond lattice as in the cubic case. The next-nearest neighbor interaction, however, splits from a face-centered cubic connectivity into distinct $J_2$ and $J_2^\prime$ interactions that define body-centered and square networks, respectively. In turn, the third-neighbor interaction $J_3$ forms a diamond lattice. This model yields a large parameter space; we defer the study of its mean-field phase diagram and role of quantum fluctuations to Sec.~\ref{sec:theory}. With the propagation vector $\mathbf{k}_{\rm m}$ and the inelastic spectrum as a constraints, we obtain an excellent match between the data and the calculated scattering intensity for $J_1=10.1$~meV, $J_2/J_1=0.32$, $J_2^\prime/J_1=0.18$ and $J_3/J_1=0.098$  [Fig.~\ref{fig:cuinelastic}]. As we will see below, this set of parameters is uniquely constrained by the experimental data. We note that in the cubic case, the average value $(J_2+J_2^\prime)/2J_1=0.125$ would yield a highly degenerate coplanar spiral state.~\cite{bergman2007order} The Jahn-Teller distortion in \curho\ is thus crucial to stabilize a well-defined spin-helix with a unique propagation vector $\mathbf{k}_{\rm m}=(0,0,0.79)$. In a trend already observed for \corho, the temperature dependence of the magnetic excitations of \curho\ [Fig.~\ref{fig:cutdep}] is marked by a very rapid collapse of the magnetic bandwidth as $T_{\rm N}$ is crossed.

%================  ================  ================  ================
\section{Theoretical analysis}
\label{sec:theory}
%================  ================  ================  ================

%------------------------------------------
\subsection{Mean-field phase diagram}
%------------------------------------------

In this section we apply mean-field theory to relate the magnetic structure of {\curho} to a Heisenberg Hamiltonian with the exchange interactions of Fig.~\ref{fig:cumag}(a). Calculations are efficiently performed in a primitive unit cell, which is less symmetric than the conventional cell but contains the smallest possible number of atoms; see Appendix~\ref{ap:primitive}. We proceed with the Heisenberg model, 
\begin{equation}
\mathcal{H}=\frac{1}{2}\sum_{i,j,m,n}J_{ij}(\mathbf{R}_{n}-\mathbf{R}_{m})\, \mathbf{S}_{i}(\mathbf{R}_{m})\cdot\mathbf{S}_{j}(\mathbf{R}_{n}),\label{eq:heisenberg_model}
\end{equation}
where $\mathbf{S}_{i}(\mathbf{R}_{m})$ denotes the $i$-th spin of a primitive unit cell located at a lattice vector $\mathbf{R}_{m}$ from the origin, and $J_{ij}(\mathbf{R}_{n}-\mathbf{R}_{m})\equiv J_{d}$ is the exchange interaction between spins $\mathbf{S}_{i}(\mathbf{R}_{m})$ and $\mathbf{S}_{j}(\mathbf{R}_{n})$. We consider the four exchange interactions $J_1$, $J_2$, $J_2^\prime$ and $J_3$ shown in Fig.~\ref{fig:cumag}(a) and neglect possible exchange anisotropies.

Our mean-field theory follows the steps of Bertaut~\cite{Bertaut_1962} and Chapon~\cite{Chapon_2009} and proceeds by
taking the Fourier transform of the exchange interactions,
\begin{equation}
	J_{ij}(\mathbf{q})=\sum_{n}J_{ij}(\mathbf{R}_{n})\exp\left(-\mathrm{i}\mathbf{q}\cdot\mathbf{R}_{n}\right),
\end{equation}
where $i\in\left\{ 1,2\right\} $ labels the two Cu ions in the primitive unit cell. $J_{ij}(\mathbf{q})$ describes a $2\times2$ Hermitian matrix for each momentum $\mathbf{q}$ in the first Brillouin zone,
\begin{equation}
	J(\mathbf{q})=-\left(\begin{array}{cc}
	J_{11} & J_{12}\\
	J_{12}^{*} & J_{11}
	\end{array}\right),
\end{equation}
where $J_{ij}^{*}(\mathbf{q})=J_{ij}(-\mathbf{q})$. The matrix elements are evaluated by identifying the lattice translation vectors that connect pairs of spins dressed by a given interaction. Using Eq.~(\ref{eq:index_convert}) to convert from primitive to conventional indices, we obtain
\begin{eqnarray}
	J_{11}(\mathbf{q}) & = & 2J_{2} \{ \cos\left[\pi\left(h+k+l\right)\right]+\cos\left[\pi\left(h+k-l\right)\right] \nonumber \\
	& & ~~~~~ + \cos\left[\pi(k+l-h)\right]+\cos\left[\pi\left(h+l-k\right)\right]\} \nonumber \\
	&+& 2J'_{2}\left[\cos\left(2\pi h\right)+\cos\left(2\pi k\right)\right],\label{eq:J_11}\\
	J_{12}(\mathbf{q}) & = & J_{1} \large\{ 1 + e^{2\pi ik} + e^{\pi i\left(h+k+l\right)}
	 + e^{\pi i\left(-h+k+l\right)}\large\} \\
	&  +& J_{3} \large\{ e^{-\pi i\left(h-k+l\right)} + e^{\pi i \left(h+k-l\right)} 
	+ e^{2\pi il} + e^{2\pi i\left(k+l\right)}\large\}, \nonumber \label{eq:J_12}
\end{eqnarray}
where $(h,k,l)$ are expressed in reciprocal lattice units of the conventional unit cell.

The interaction matrix has two eigenvalues at each wavevector $\mathbf{q}$, given
by
\begin{equation}
\lambda_{\pm}(\mathbf{q})=J_{11}(\mathbf{q})\pm\left|J_{12}(\mathbf{q})\right|.\label{eq:eigenvalues}
\end{equation}
The wavevector $\mathbf{k}$ for which $\max\left[\lambda_{\pm}(\mathbf{k})\right]$ reaches a global maximum in the first Brillouin zone is associated with the propagation vector $\mathbf{k}_{\rm m}$ of the ordered magnetic state. Only a small number of $\mathbf{k}$ points related by symmetry usually fulfill this condition. Highly-frustrated systems are exceptions for which $\max\left[\lambda(\mathbf{q})\right]$ can be degenerate over large regions of the Brillouin zone.~\cite{Reimers_1991} Given the large parameter space, a systematic search for maximum eigenvalues as a function of $J_1$, $J_2$, $J_2^\prime$ and $J_3$ is very time consuming. Minimization of the classical ground-state energy can significantly reduce the computing burden by providing analytical solutions for the magnetic structure, see Appendix~\ref{ap:gs}.

\begin{figure}[t!]
	\includegraphics[width=0.59\columnwidth]{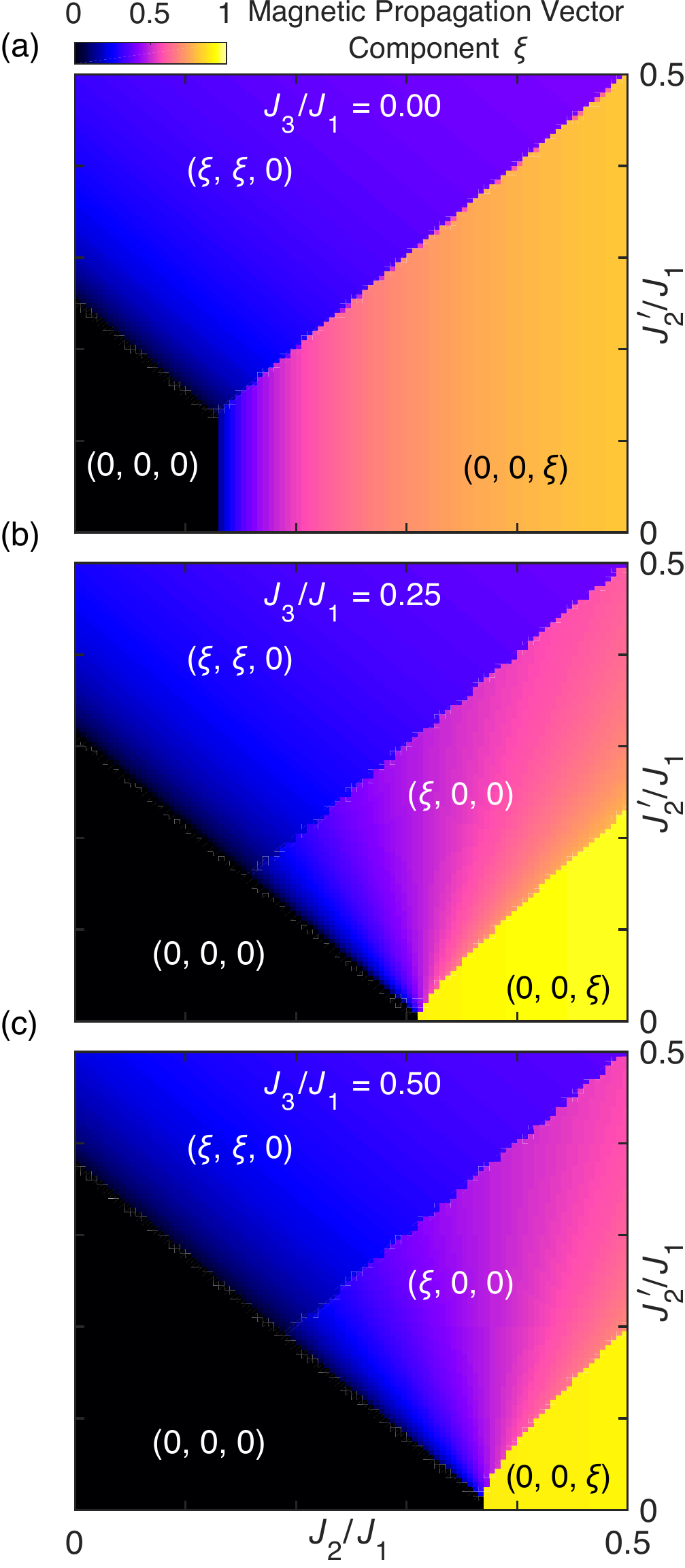} 
	\caption{Mean-field phase diagrams for the {\curho} Hamiltonian of Fig.~\ref{fig:cumag}(a) with different $J_2/J_1$, $J_2^\prime/J_1$ and $J_3/J_1$. The color represents the free components of the magnetic propagation vector in the different phases we uncover.}
	\label{fig:phasediagram}
\end{figure}

Our mean-field phase diagram as a function of $J_2/J_1$ and $J_2^\prime/J_1$ for different values of $J_3$ and assuming all exchanges antiferromagnetic is shown in Fig.~\ref{fig:phasediagram}. As well as of the N\'eel phase, we identify three different incommensurate phases for which the magnetic propagation vector takes the form $\mathbf{k}_{{\rm m},1}\!=\!(\xi, \xi, 0)$, $\mathbf{k}_{{\rm m},2}\!=\!(\xi, 0, 0)$ (which is equivalent to $(0, \xi, 0)$) or $\mathbf{k}_{{\rm m},3}\!=\!(0, 0, \xi)$. The propagation vector observed for \curho, $\mathbf{k}_{{\rm m},3}$, is stabilized with an incommensurate $\xi$ value for a broad range of $J_2/J_1$ and $J_2^\prime/J_1$ values. A large $J_3/J_1$, however, pins the spiral to the lattice and leads to $\mathbf{k}_{{\rm m},3}\!=\!(0, 0,1)$. Critically, our results indicate that the value of $\mathbf{k}_{{\rm m},3}$ is only affected by $J_2/J_1$ and $J_3/J_1$ but not by $J^\prime_2/J_1$. Therefore, the experimentally-measured value of the propagation vector constraints the ratio of $J_3/J_1$ to $J_2/J_1$, which eliminates one degree of freedom when simulating the excitations of {\curho} with linear spin wave theory.

%================
%================

%------------------------------------------
\subsection{Linear spin-wave theory}
%------------------------------------------

With the knowledge of the possible magnetic structures of the model, we resort to linear spin-wave theory to simulate the dynamics of spins in both coumpounds and to refine further the exchanges parameters for \curho\ [Fig.~\ref{fig:cumag}(a)]. For {\corho} we only consider the nearest-neighbor coupling. While the simulated scattering intensities of Figs.~\ref{fig:coinelastic} and \ref{fig:cuinelastic} are obtained numerically using SpinW,~\citep{Toth_2015} we proceed below with the explicit calculation of the magnon dispersion, a step necessary to calculate the effect of zero-point quantum fluctuations on the magnetic ordering.

We start with the general case of \curho\ for which we set the spins to lie in the $x_0$-$y_0$ plane of the laboratory reference frame (the conventional unit cell). In order to align the quantization axis $z$ along the direction of each spin, we perform the following transformation, \begin{gather}
S_{i}^{x_0}(\mathbf{R}_{n}) = -S_{i}^{y}(\mathbf{R}_{n}) \sin\theta + S_{i}^{z}(\mathbf{R}_{n}) \cos\theta, \\
S_{i}^{y_0}(\mathbf{R}_{n}) = -S_{i}^{y}(\mathbf{R}_{n}) \cos\theta - S_{i}^{z}(\mathbf{R}_{n}) \sin\theta, \\
S_{i}^{z_0}(\mathbf{R}_{n}) = S_{i}^{x}(\mathbf{R}_{n}),
\end{gather}
where $\theta$ is related to the propagation vector $\mathbf{k}_{\rm m}$ and is a function of $i$ and $\mathbf{R}_{n}$. We then introduce the Holstein-Primakoff $a$-bosons, which to linear order relate to spin operators in the rotating frame as
\begin{eqnarray}
	S_{i}^z(\mathbf{R}_{n}) &=& S - a_{i,\mathbf{R}_{n}}^{\dagger}a_{i,\mathbf{R}_{n}}, \\
	\label{eq:ordmom}
	S_{i}^x(\mathbf{R}_{n}) &\approx& \sqrt{S/2}(a_{i,\mathbf{R}_{n}} + a_{i,\mathbf{R}_{n}}^{\dagger}), \\
	S_{i}^y(\mathbf{R}_{n}) &\approx& -i\sqrt{S/2}(a_{i,\mathbf{R}_{n}} - a_{i,\mathbf{R}_{n}}^{\dagger}),
\end{eqnarray}
and Fourier transform as
\begin{equation}
	a_{i,\mathbf{R}_{n}} = \sum\limits_{\mathbf{q}\in\mathrm{B.Z.}} a_{i, \mathbf{q}} e^{i\mathbf{q}\cdot\mathbf{R}_{n}}.
\end{equation}

Keeping only quadratic terms in boson operators, we obtain the Hamiltonian
\begin{equation}
	\hat{\mathcal{H}}_2 = \frac{1}{2} \sum\limits_{\mathbf{q}\in\mathrm{B.Z.}} \boldsymbol{X}_{\mathbf{q}}^{\dagger} H_{\mathbf{q}} \boldsymbol{X}_{\mathbf{q}},
\end{equation}
where $\boldsymbol{X}_{\mathbf{q}}^{\dagger} \equiv (a_{1,\mathbf{q}}^\dagger, a_{2,\mathbf{q}}^\dagger, a_{1, -\mathbf{q}}, a_{2, -\mathbf{q}})$ is a row vector of boson operators and $\boldsymbol{X}_{\mathbf{q}}$ the corresponding column vector.  In this representation, $H_\mathbf{q}$ is a $4\times 4$ Hermitian matrix which can be  diagonalized provided the following bosonic commutation rules are preserved:
\begin{equation}
	g = \boldsymbol{X}_{\mathbf{q}}\boldsymbol{X}_{\mathbf{q}}^{\dagger} - \boldsymbol{X}_{\mathbf{q}}^{\ast}\boldsymbol{X}_{\mathbf{q}}^{\rm{T}} = \left( \begin{array}{cccc}
	1 & 0 & 0 & 0\\
	0 & 1 & 0 & 0\\
	0 & 0 & -1 & 0\\
	0 & 0 & 0 & -1\end{array} \right).
\end{equation}
For the $\mathbf{k}_{m,3}=(0,0,\xi)$ magnetic structure observed in \curho, the matrix elements $\{h_{ij}\}$ of $H_\mathbf{q}$ read:
\begin{align}
	h_{11} & = h_{22} = h_{33} = h_{44} = \\\nonumber
	& -4S\big[J_1\cos\varphi + 2J_2\cos\gamma + J_2' - J_2'\zeta_{2^\prime}(\mathbf{q})\\\nonumber
	& - J_2(1 + \cos\gamma)\zeta_{2}(\mathbf{q}) + J_3\cos3\varphi\big], \\
	h_{12} & = h_{34} = \\\nonumber
	& 2S\big[J_1(1+\cos\varphi)\zeta_1(\mathbf{q}) + J_3(1+\cos3\varphi)\zeta_{3}(\mathbf{q})\big], \\
	h_{13} & = h_{31} = h_{24} = h_{42} = \\\nonumber
	& 4SJ_2(1 -\cos\gamma)\zeta_{2}(\mathbf{q}), \\
	h_{14} & = h_{32} = \\\nonumber
	& 2S\big[J_1(1-\cos\varphi)\zeta_{1}(\mathbf{q}) + J_3(1-\cos3\varphi)\zeta_{3}(\mathbf{q})\big], \\
	h_{21} & = h_{43} = \\\nonumber
	& 2S\big[J_1(1+\cos\varphi)\zeta^\ast_{1}(\mathbf{q}) + J_3(1+\cos3\varphi)\zeta^\ast_{3}(\mathbf{q})\big], \\
	h_{23} & = h_{41} = \\\nonumber
	& 2S\big[J_1(1-\cos\varphi)\zeta^\ast_{1}(\mathbf{q}) + J_3(1-\cos3\varphi)\zeta^\ast_{3}(\mathbf{q})\big], 
\end{align}
where $\zeta_i(\mathbf{q})$'s are the lattice harmonics associated with exchange $J_i$,
\begin{align}
	\zeta_{1}(\mathbf{q}) & = \\\nonumber 
	& \frac{1}{4}\left(1+e^{-i\pi(-h+k+l)}+e^{-i2\pi k}+e^{-i\pi(h+k+l)}\right),\\
	\zeta_{2^\prime}(\mathbf{q}) & = \frac{1}{2}\Big(\cos[2\pi k]+\cos[2\pi h]\Big),\\
	\zeta_{2}(\mathbf{q}) & =  \frac{1}{4}\Big(\cos[\pi(-h+k+l)]+\cos[\pi(h-k+l)] \\\nonumber
	&~~~~~+\cos[\pi(h+k-l)]+\cos[\pi(h+k+l)]\Big),\\
	\zeta_{3}(\mathbf{q}) & = \\\nonumber
	& \frac{1}{4}\Big(e^{i\pi(h-k+l)}+e^{-i\pi(h+k-l)}+e^{-2i\pi l}+e^{-2i\pi(k+l)}\Big),
\end{align}
with $\mathbf{q} = (h, k, l)$ in reciprocal lattice units of the conventional unit cell.

In general, it is not possible to give an analytical form for the above eigenvalue problem. We thus follow the numerical solution described by S. Petit.~\cite{Petit_2011} First, we perform a Cholesky decomposition on $H_{\mathbf{q}}$ to find $K_\mathbf{q}$ that satisfies $H_\mathbf{q} = K^\dagger_\mathbf{q} K_\mathbf{q}$. The positive definiteness for $H_\mathbf{q}$ is guaranteed provided the ground state minimizes the classical energy. Afterwards we numerically diagonalize $K_\mathbf{q} g K^\dagger_\mathbf{q}$. The eigenvalues of the resulting diagonal matrix $gD_\mathbf{q}$ provide the magnon energies $\pm\omega_i(\mathbf{q})$ ($i = 1, 2$). To obtain the eigenvectors, we sort the positive eigenvalues in ascending order and sort the corresponding negative ones accordingly. The transformation matrix $V_\mathbf{q}$ that leads to new boson operators $\{b_{i, \mathbf{q}}\}$ from $\{a_{i, \mathbf{q}}\}$ bosons is calculated in the following way
\begin{equation}
	V_\mathbf{q} = K_\mathbf{q}^{-1} U_\mathbf{q} D_\mathbf{q}^{1/2},
\label{eq:transform}
\end{equation}
where the unitary transformation matrix $U_\mathbf{q}$ makes $K_\mathbf{q} g K^\dagger_\mathbf{q}$ diagonal. Note that $V_\mathbf{q}$ is not unitary and it is normalized through $V_\mathbf{q}^\dagger g V_\mathbf{q} = g$.

In the case of \corho, the N{\'e}el ground state allows to write an explicit analytical solution for the magnon energies. We can explicitly write down the quadratic Hamiltonian as
\begin{equation}
	\hat{\mathcal{H}}_2 = 4J_1S\sum\limits_{i,j,\mathbf{q}} \left[\delta_{ij} a_{i,\mathbf{q}}^\dagger a_{j,\mathbf{q}} - \frac{1}{2}\Lambda^{ij}_{\mathbf{q}}(a_{i,\mathbf{q}}^\dagger a_{j,-\mathbf{q}}^\dagger + \mathrm{h.c.}) \right],
\end{equation}
such that the matrix $H_{\mathbf{q}}$ reads
\begin{equation}
	H_{\mathbf{q}} = 4J_1S\left( \begin{array}{cc}
	A_{\mathbf{q}} & B_{\mathbf{q}} \\
	B_{\mathbf{q}} & A_{\mathbf{q}} \end{array} \right),
\end{equation}
with $A_{\mathbf{q}} = I_2$ the $2\times2$ identity matrix, and
\begin{equation}
	B_{\mathbf{q}} = \frac{1}{4} \left( \begin{array}{cc}
	0 & \Lambda_{\mathbf{q}} \\
	\Lambda_{\mathbf{q}}^\ast & 0 \end{array} \right),
\end{equation}
with $\Lambda_{\mathbf{q}} = 1 + e^{i\pi(h+k)} + e^{i\pi(h+l)} + e^{i\pi(k+l)}$.

From here, the calculation proceeds as for {\curho}, or alternatively a ``two-step diagonalization" \cite{PhysRevB.92.144415} can be applied due to the evident commutativity of $A_{\mathbf{q}}$ and $B_{\mathbf{q}}$. We first apply the unitary transformation
\begin{equation}
	a_{i,\mathbf{q}}=\sum\limits_{j}w_{j i,\mathbf{q}}d_{j,\mathbf{q}}
\label{eq:unitary}
\end{equation}
to rewrite the quadratic Hamiltonian as
\begin{equation}
	\hat{\mathcal{H}}_2 = 4J_1S\sum\limits_{i,\mathbf{q}} \left[d_{i,\mathbf{q}}^\dagger d_{i,\mathbf{q}} - \frac{1}{2}\lambda_{i,\mathbf{q}}(d_{i,\mathbf{q}}^\dagger d_{i,-\mathbf{q}}^\dagger + \mathrm{h.c.}) \right],
\end{equation}
where $\lambda_{i,\mathbf{q}} = \pm \left|\Lambda_{\bf q}\right|$ are the eigenvalues of $B_{\mathbf{q}}$. This eliminates the cross terms between two types of boson operators and effectively leaves two independent single-boson Hamiltonians. From there, we perform the conventional Bogolyubov transformation for each individual species of $d$-bosons,
\begin{equation}
	d_{i,\mathbf{q}} = u_{i, \mathbf{q}}b_{i,\mathbf{q}} + v_{i, \mathbf{q}}b_{i,-\mathbf{q}}^\dagger ,
\label{eq:bogo}
\end{equation}
under the constraint $u_{i, \mathbf{q}}^2 - v_{i, \mathbf{q}}^2 = 1$.  The solution for $u_{i \mathbf{q}}^2$ and $v_{i \mathbf{q}}^2$ is
\begin{align}
	u_{i, \mathbf{q}}^2,v_{i, \mathbf{q}}^2 & = \frac{1}{2}\left(\frac{1}{\omega_\mathbf{q}} \pm 1\right), \\
	2u_{i, \mathbf{q}}v_{i, \mathbf{q}} & = \frac{\lambda_{i,\mathbf{q}}}{\omega_\mathbf{q}},
\end{align}
where
\begin{equation}
	\omega_\mathbf{q} = 4J_1S\sqrt{1-\frac{\left|\Lambda_{\bf q}\right|^2}{16}}
\end{equation}
is the two-fold degenerate dispersion relation for \corho.

\subsection{Zero-point spin reduction}

%================
\begin{figure}[h!]
	\includegraphics[width=0.90\columnwidth]{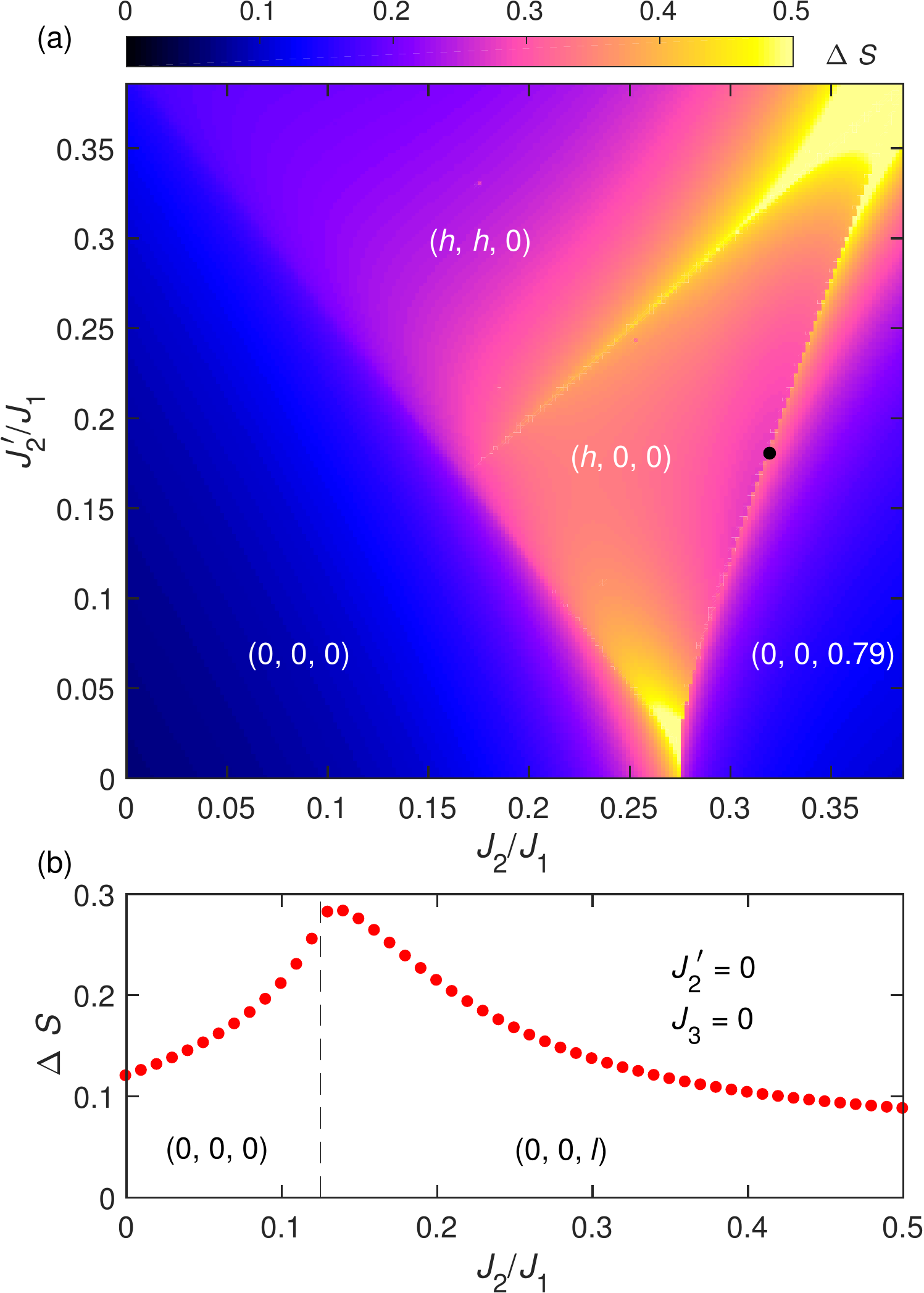} 
	\caption{{Zero-point spin reduction calculation for \curho. (a) $\Delta S$ as a function of $J_2/J_1$ and $J_2^\prime/J_1$ with $J_3$ varying in order to maintain the propagation vector $\mathbf{k}_{\rm m}=(0, 0, 0.79)$ spiral phase whenever possible. The black dot corresponds to the parameters obtained for {\curho}. (b) $\Delta S$ as a function of $J_2/J_1$ for $J_2^\prime$ and $J_3$ fixed to zero. The vertical dashed line is located at the classical transition ratio ($J_2/J_1 = 1/8$) from the N{\'e}el state to the incommensurate spiral state.}}
	\label{fig:quantum}
\end{figure}
%================

To evaluate the strength of quantum effects in our diamond-lattice antiferromagnets, we calculate the $1/S$ zero-point reduction on the ordered moment due to quantum fluctuations. In general, the spin reduction $\Delta S_{i}$ is sub-lattice dependent and reads
\begin{equation}
	\Delta S_{i} \equiv S - \left< S_{i,\mathbf{R}_{n}}^z \right> = \left< a_{i,\mathbf{R}_{n}}^\dagger a_{i,\mathbf{R}_{n}} \right> = \frac{1}{\mathcal{N}} \sum\limits_{\mathbf{k}\in\mathrm{B.Z.}} \left< a_{i,\mathbf{q}}^\dagger a_{i,\mathbf{q}} \right>,
\end{equation}
where $\left< \cdots \right>$ is the thermal average and $\mathcal{N}$ is the number of unit cells in the entire system. Our approach to evaluate $\Delta S_{i}$ uses the transformation matrix $V_\mathbf{q}$ obtained from Eq.~(\ref{eq:transform}) to transform $a$-bosons into $b$-bosons. As a result,
\begin{align}
\left< a_{i,\mathbf{q}}^\dagger a_{i,\mathbf{q}} \right> = & \sum\limits_{j=1}^2 V^\dagger_{j+2, i} V_{i, j+2} \\\nonumber
& + \left[V^\dagger_{j i} V_{i j} + V^\dagger_{j+2, i} V_{i, j+2}\right]\left< b_{j,\mathbf{q}}^\dagger b_{j,\mathbf{q}} \right>,
\end{align}
where boson commutation relations are applied. As $\left< b_{i, \mathbf{q}}^\dagger b_{i, \mathbf{q}} \right>$ vanishes in the limit $T\!=\!0$, this yields the general formula for the zero-point spin reduction
\begin{align}
\label{eq:sprd1}
\Delta S_{i} & = \frac{1}{\mathcal{N}} \sum\limits_{\mathbf{q}\in\mathrm{B.Z.}}\sum\limits_{j=1}^2 V^\dagger_{j+2, i} V_{i, j+2} \\\nonumber
& = \int_{\mathbf{q}\in\mathrm{B.Z.}} \mathrm{d}^3 {\bf q}\sum\limits_{j=1}^2 V^\dagger_{j+2, i} V_{i, j+2} \ (\mathcal{N} \rightarrow \infty).
\label{eq:sprd1}
\end{align}
When the ``two-step diagonalization" is applicable, Eqns.~(\ref{eq:unitary}, \ref{eq:bogo}) allow us to rewrite the spin reduction in the more traditional form
\begin{equation}
\Delta S_{i} = \frac{1}{\mathcal{N}} \sum\limits_{\mathbf{q}\in\mathrm{B.Z.}}\sum\limits_{j=1}^2 w^2_{i j,\mathbf{q}}v^2_{j,\mathbf{q}},
\end{equation}
which can be further simplified assuming the two magnetic sites experience the same zero-point reduction,
\begin{align}
\label{eq:sprd2}
\Delta S & = \frac{1}{2}\sum\limits_{j=1}^2\int_{\mathbf{q}\in\mathrm{B.Z.}} v^2_{j,\mathbf{q}} \mathrm{d}^3 q  \ (\mathcal{N} \rightarrow \infty).
\end{align}

 The numerical calculation of Eq.~\ref{eq:sprd1} was implemented in {\CC} with the help of the adaptive multidimensional integration algorithm.\cite{Genz1980, Berntsen1991} Estimated integration errors are generally under 0.5\% except for some of critical values of $J's$, {\eg}, $J_2 = J_2'$, for which integration errors are slightly larger. Setting $J_2$, $J_2^\prime$ and $J_3$ to zero yields the spin reduction value for the nearest-neighbor 3D diamond-lattice antiferromagnet, $\Delta S_{\Diamond}\!=\!0.11973(1)$, a number significantly smaller than for the nearest-neighbor 2D square-lattice antiferromagnet, $\Delta S_{\square}\!=\! 0.19660$.~\cite{Igarashi_1992} 

 When competing exchanges relevant for \curho\ are included, however, we find that the $1/S$ zero-point reduction dramatically increases for exchange parameters in the vicinity of the mean-field transition lines; see Fig.~\ref{fig:quantum}. For the exchange parameters of {\curho}, we obtain a spin reduction of $\Delta S_{\rm Cu}\!=\!0.330(1)$ and thus predict an ordered moment of $\langle \mu \rangle = g (S - \Delta S_{\rm Cu}) \mu_{\rm B} \approx 0.36 \mu_{\rm B}$, in qualitative agreement with the strongly reduced moment $\mu_{\rm ord}\!=\! 0.56(6) \mu_{\rm B}$ obtained experimentally. We conclude that such a strong reduction of the ordered moment for a 3D magnet originates from competing exchange interactions that place {\curho} in the vicinity of a transition line between $\mathbf{k}_{{\rm m},3}=(0,0,\xi)$ and $\mathbf{k}_{{\rm m},2}=(\xi,0,0)$ orders.

%================  ================  ================  ================
\section{Conclusion}
\label{sec:concl}
%================  ================  ================  ================

In conclusion, our detailed experimental and theoretical work identifies the $A$-site spinels \corho\ and \curho\ as model diamond-lattice antiferromagnets. The cubic compound  \corho\ is a canonical realization of the nearest-neighbor Heisenberg antiferromagnet on the diamond lattice. Below the N\'eel temperature $T_{\rm N}\!\approx\!25$~K, the $S=3/2$ magnetic moments order in a bipartite antiferromagnetic structure with a negligible contribution from zero-point fluctuations. Neutron scattering experiments for $T\ll T_{\rm N}$ reveal well-formed spin-wave excitations, successfully described by a single nearest-neighbor magnetic exchange parameter $J_1=6.3$~meV. Around and above $T_{\rm N}$, the magnetic excitations rapidly soften and become strongly damped, as expected for a three-dimensional antiferromagnet.

In the tetragonally-distorted $S=1/2$ spinel \curho, the magnetism is much richer and strongly influenced by the competition between nearest-neighbor and further-neighbor exchange interactions. While thermodynamic probes paint a picture remarkably similar to \corho, with a N\'eel temperature of $T_{\rm N}\approx24$~K, neutron scattering reveals that the magnetic structure of \curho\ is an incommensurate spin helix with a propagation vector $\mathbf{k}_{\rm m}=(0,0,0.79)$ and strong quantum reduction of the ordered magnetic moments to $\approx50$\% of their classical value. Comparison between inelastic neutron scattering results and spin-wave theory provides a quantitative understanding of the underlying microscopic mechanism responsible for this unexpected ground state. Due to the tetragonal lattice symmetry, the degeneracy between second-neighbor exchanges ($J_2$ and $J_2^\prime$) is lifted when compared to the cubic case. Our mean-field calculation shows that the competition of these exchanges with first- ($J_1$) and third-neighbor ($J_3$) interactions stabilizes the incommensurate spin helix observed experimentally. Remarkably, we find that \curho\ lies close to a transition between two distinct magnetically ordered ground-states. Using $1/S$-corrections to the ordered moment, we find that zero-point fluctuations are enhanced for the exchange parameters of \curho, which explains the strong moment reduction observed experimentally. 

Overall, our results add two model magnets to the expanding family of diamond-lattice antiferromagnets and further demonstrate the importance of competing exchange interactions and possible enhancement of quantum effects in such systems. We expect our results on \curho\ to guide future studies to elucidate the spin-liquid phenomenology recently uncovered in the isostructural compound NiRh$_2$O$_4$,~\cite{chamorro2017frustrated} and more generally to contribute to the search for the predicted topological paramagnetism in $S=1$ diamond-lattice antiferromagnets.~\cite{Senthil_2015} On the methodological side, our work demonstrates that combining state-of-the-art neutron scattering experiments with mean-field and spin-wave theory modeling allows to extract definitive microscopic information from polycrystalline samples alone, even when magnetic correlations are three-dimensional. This is important to accelerate the search for exotic quantum states in real systems through the screening of many related materials, an endeavor that would be too costly, difficult or slow to undertake on single-crystalline samples.

%======================================================================
% Acknowledgments
%======================================================================
\begin{acknowledgments} 
	The authors thank T.~M.~McQueen for discussion and T.~Senthil for motivating this work. The work at Georgia Tech (L.G., J.A.M.P, M.M.) was supported by the College of Sciences and Oak Ridge Associated Universities through a Ralph E. Powe Junior Faculty Enhancement Award (M.M.). The work at Oregon State University (M.A.S) was supported by the National Science Foundation through grant DMR-1534711. The work at UC Santa Cruz (A.P.R.) was supported by the National Science Foundation through grant DMR-1534741. The research at Oak Ridge National Laboratory's Spallation Neutron Source and High Flux Isotope Reactor was sponsored by the U.S. Department of Energy, Office of Basic Energy Sciences, Scientific User Facilities Division. We are grateful to A. Huq for collecting data through the mail-in program at ORNL's POWGEN.
\end{acknowledgments}

%======================================================================
% Appendix
%======================================================================
\appendix
\section{Conventional and primitive cells}
\label{ap:primitive}

In the case of \curho, the primitive $(\boldsymbol{a}_{\rm p},\boldsymbol{b}_{\rm p},\boldsymbol{c}_{\rm p})$ and conventional $(\boldsymbol{a},\boldsymbol{b},\boldsymbol{c})$ crystallographic unit cells are related by:
\begin{equation}
	\left( \begin{array}{c} 
		\boldsymbol{a}_{\rm p} \\ \boldsymbol{b}_{\rm p} \\ \boldsymbol{c}_{\rm p}
	\end{array} \right) 
	\!=\! \frac{1}{2}
	\left(\begin{array}{rrr}	   
		1 & 1 & -1\\
		-1 & 1 & 1 \\
		1 & -1 & 1 
	\end{array} \right)
	\left( \begin{array}{c} 
		\boldsymbol{a} \\ \boldsymbol{b} \\ \boldsymbol{c}
	\end{array} \right).
\end{equation}
This yields the following relations between the Miller indices $(h,k,l)$
and atomic fractional coordinates $(x,y,z)$ for the two unit cells,
\begin{eqnarray}
	\left(x_{\mathrm{p}},y_{\mathrm{p}},z_{\mathrm{p}}\right) & = & \left(x+y,y+z,x+z\right) \\
	\left(2h_{\mathrm{p}},2k_{\mathrm{p}},2l_{\mathrm{p}}\right) & = & \left(h+k+l,-h+k+l,h-k+l\right),  \nonumber
	\label{eq:index_convert}
\end{eqnarray}
such that the primitive unit cell of \curho\ contains two Cu atoms,
at fractional coordinates $\mathbf{r}_{1}=(0,0,0)$ and $\mathbf{r}_{2}=(\frac{1}{2},\frac{3}{4},\frac{1}{4})$.

\section{Classical ground state energy minimization}
\label{ap:gs}

In this appendix, we provide general expressions for the classical ground-state energy per spin and magnetic structure for different phases of our mean-field phase diagram. Working in the primitive unit cell, the magnetic structure is fully defined from the knowledge of $\varphi$, the angle between two spins in one primitive cell, and $\gamma$ the pitch angle between neighbor cells that enters the propagation vector $\mathbf{k}_{\rm m}$. 

In the case of the $\mathbf{k}_{{\rm m},3} = (0,0,\gamma/\pi)$ phase, the classical energy per spin reads
\begin{align*}
\mathcal{H}_0 = S^2\left\{\right.&\left.J_1\left[\cos\varphi + \cos(\varphi - \gamma )\right] + 4J_2\cos\gamma + 2J_2' \right. \\
+ & \left. J_3\left[\cos(2\gamma-\varphi)+\cos(\gamma+\varphi)\right]\right\}.
\end{align*}
The local minimum conditions, ${\partial\mathcal{H}_0}/{\partial\varphi} = 0$ and ${\partial\mathcal{H}_0}/{\partial\gamma} = 0$ and the intuitive assumption of a uniform angle between spins along $\boldsymbol{c}$, $\cos\gamma = \cos2\varphi$, lead to
\begin{eqnarray*}
	\cos\varphi &=& -\frac{1}{6J_3}\left(\sqrt{4J_2^2-3J_1J_3+9J_3^2}-2J_2\right) \ (J_3 \neq 0), \\
	\cos\varphi &=& -\frac{J_1}{8J_2} \ (J_3 = 0).
\end{eqnarray*}

Similarly, the expression for the classical energy per spin in the $\mathbf{k}_{{\rm m},2} = (\gamma/\pi,0,0)$ phase is
\begin{align*}
\mathcal{H}_0 = S^2\left\{\right. & (J_1 + J_3)\left[2\cos\varphi + \cos(\varphi - 2\gamma) + \cos(\varphi + 2\gamma)\right] \\
+ & \left. 8J_2\cos\gamma + 2J_2^\prime\cos 2\gamma \right\}.
\end{align*}
This is not easy to solve directly without any extra information. However, with the help of mean-field calculations, we find $\varphi = \pi$, which yields
\begin{equation*}
\cos\gamma = \frac{J_1+J_3-4J_2}{4J_2^\prime}.
\end{equation*}

Finally, for the $\mathbf{k}_{{\rm m},1} = (\gamma/\pi,\gamma/\pi,0)$ phase, the expression for the classical energy per spin is
\begin{align*}
\mathcal{H}_0 = S^2\left\{\right. & \left.(J_1 + J_3)\left[\cos\varphi + \cos(\varphi - \gamma )\right] \right.\\
+ & \left.2J_2\cos\gamma + 2J_2'\cos\gamma + 2J_2\right\}.
\end{align*}
It is straightforward to solve the local minimum conditions without making any assumptions, which yields
\begin{align*}
	\cos\varphi = -\frac{J_1+J_3}{4(J_2+J_2')}; \\
	\cos\gamma = -1 + \frac{(J_1+J_3)^2}{8(J_2+J_2')^2}.
\end{align*}

%======================================================================
% Bibliography
%======================================================================
%\bibliographystyle{apsrev4-1}
%\bibliography{cucorho}

\begin{thebibliography}{65}%
\makeatletter
\providecommand \@ifxundefined [1]{%
 \@ifx{#1\undefined}
}%
\providecommand \@ifnum [1]{%
 \ifnum #1\expandafter \@firstoftwo
 \else \expandafter \@secondoftwo
 \fi
}%
\providecommand \@ifx [1]{%
 \ifx #1\expandafter \@firstoftwo
 \else \expandafter \@secondoftwo
 \fi
}%
\providecommand \natexlab [1]{#1}%
\providecommand \enquote  [1]{``#1''}%
\providecommand \bibnamefont  [1]{#1}%
\providecommand \bibfnamefont [1]{#1}%
\providecommand \citenamefont [1]{#1}%
\providecommand \href@noop [0]{\@secondoftwo}%
\providecommand \href [0]{\begingroup \@sanitize@url \@href}%
\providecommand \@href[1]{\@@startlink{#1}\@@href}%
\providecommand \@@href[1]{\endgroup#1\@@endlink}%
\providecommand \@sanitize@url [0]{\catcode `\\12\catcode `\$12\catcode
  `\&12\catcode `\#12\catcode `\^12\catcode `\_12\catcode `\%12\relax}%
\providecommand \@@startlink[1]{}%
\providecommand \@@endlink[0]{}%
\providecommand \url  [0]{\begingroup\@sanitize@url \@url }%
\providecommand \@url [1]{\endgroup\@href {#1}{\urlprefix }}%
\providecommand \urlprefix  [0]{URL }%
\providecommand \Eprint [0]{\href }%
\providecommand \doibase [0]{http://dx.doi.org/}%
\providecommand \selectlanguage [0]{\@gobble}%
\providecommand \bibinfo  [0]{\@secondoftwo}%
\providecommand \bibfield  [0]{\@secondoftwo}%
\providecommand \translation [1]{[#1]}%
\providecommand \BibitemOpen [0]{}%
\providecommand \bibitemStop [0]{}%
\providecommand \bibitemNoStop [0]{.\EOS\space}%
\providecommand \EOS [0]{\spacefactor3000\relax}%
\providecommand \BibitemShut  [1]{\csname bibitem#1\endcsname}%
\let\auto@bib@innerbib\@empty
%</preamble>
\bibitem [{\citenamefont {Affleck}(1989)}]{Affleck_1989}%
  \BibitemOpen
  \bibfield  {author} {\bibinfo {author} {\bibfnamefont {I.}~\bibnamefont
  {Affleck}},\ }\href {http://stacks.iop.org/0953-8984/1/i=19/a=001} {\bibfield
   {journal} {\bibinfo  {journal} {J. Phys. Condens. Matter}\ }\textbf
  {\bibinfo {volume} {1}},\ \bibinfo {pages} {3047} (\bibinfo {year}
  {1989})}\BibitemShut {NoStop}%
\bibitem [{\citenamefont {Mikeska}\ and\ \citenamefont
  {Kolezhuk}(2004)}]{Mikeska_2004}%
  \BibitemOpen
  \bibfield  {author} {\bibinfo {author} {\bibfnamefont {H.-J.}\ \bibnamefont
  {Mikeska}}\ and\ \bibinfo {author} {\bibfnamefont {A.~K.}\ \bibnamefont
  {Kolezhuk}},\ }in\ \href {http://doi.org/10.1007/978-3-642-13290-2_11} {\emph
  {\bibinfo {booktitle} {Quantum magnetism}}}\ (\bibinfo  {publisher}
  {Springer},\ \bibinfo {year} {2004})\ pp.\ \bibinfo {pages}
  {1--83}\BibitemShut {NoStop}%
\bibitem [{\citenamefont {Lake}\ \emph {et~al.}(2005)\citenamefont {Lake},
  \citenamefont {Tennant}, \citenamefont {Frost},\ and\ \citenamefont
  {Nagler}}]{Lake_2005}%
  \BibitemOpen
  \bibfield  {author} {\bibinfo {author} {\bibfnamefont {B.}~\bibnamefont
  {Lake}}, \bibinfo {author} {\bibfnamefont {D.~A.}\ \bibnamefont {Tennant}},
  \bibinfo {author} {\bibfnamefont {C.~D.}\ \bibnamefont {Frost}}, \ and\
  \bibinfo {author} {\bibfnamefont {S.~E.}\ \bibnamefont {Nagler}},\ }\href
  {http://www.nature.com/nmat/journal/v4/n4/abs/nmat1327.html} {\bibfield
  {journal} {\bibinfo  {journal} {Nat. Mat.}\ }\textbf {\bibinfo {volume}
  {4}},\ \bibinfo {pages} {329} (\bibinfo {year} {2005})}\BibitemShut {NoStop}%
\bibitem [{\citenamefont {Coldea}\ \emph {et~al.}(2010)\citenamefont {Coldea},
  \citenamefont {Tennant}, \citenamefont {Wheeler}, \citenamefont {Wawrzynska},
  \citenamefont {Prabhakaran}, \citenamefont {Telling}, \citenamefont
  {Habicht}, \citenamefont {Smeibidl},\ and\ \citenamefont
  {Kiefer}}]{Coldea_2010}%
  \BibitemOpen
  \bibfield  {author} {\bibinfo {author} {\bibfnamefont {R.}~\bibnamefont
  {Coldea}}, \bibinfo {author} {\bibfnamefont {D.}~\bibnamefont {Tennant}},
  \bibinfo {author} {\bibfnamefont {E.}~\bibnamefont {Wheeler}}, \bibinfo
  {author} {\bibfnamefont {E.}~\bibnamefont {Wawrzynska}}, \bibinfo {author}
  {\bibfnamefont {D.}~\bibnamefont {Prabhakaran}}, \bibinfo {author}
  {\bibfnamefont {M.}~\bibnamefont {Telling}}, \bibinfo {author} {\bibfnamefont
  {K.}~\bibnamefont {Habicht}}, \bibinfo {author} {\bibfnamefont
  {P.}~\bibnamefont {Smeibidl}}, \ and\ \bibinfo {author} {\bibfnamefont
  {K.}~\bibnamefont {Kiefer}},\ }\href
  {http://science.sciencemag.org/content/327/5962/177} {\bibfield  {journal}
  {\bibinfo  {journal} {Science}\ }\textbf {\bibinfo {volume} {327}},\ \bibinfo
  {pages} {177} (\bibinfo {year} {2010})}\BibitemShut {NoStop}%
\bibitem [{\citenamefont {Ramirez}(1994)}]{Ramirez_1994}%
  \BibitemOpen
  \bibfield  {author} {\bibinfo {author} {\bibfnamefont {A.}~\bibnamefont
  {Ramirez}},\ }\href
  {http://www.annualreviews.org/doi/abs/10.1146/annurev.ms.24.080194.002321}
  {\bibfield  {journal} {\bibinfo  {journal} {Annu. Rev. Mater. Sci.}\ }\textbf
  {\bibinfo {volume} {24}},\ \bibinfo {pages} {453} (\bibinfo {year}
  {1994})}\BibitemShut {NoStop}%
\bibitem [{\citenamefont {Lee}(2008)}]{Lee_2008}%
  \BibitemOpen
  \bibfield  {author} {\bibinfo {author} {\bibfnamefont {P.~A.}\ \bibnamefont
  {Lee}},\ }\href {http://science.sciencemag.org/content/321/5894/1306}
  {\bibfield  {journal} {\bibinfo  {journal} {Science}\ }\textbf {\bibinfo
  {volume} {321}},\ \bibinfo {pages} {1306} (\bibinfo {year}
  {2008})}\BibitemShut {NoStop}%
\bibitem [{\citenamefont {Han}\ \emph {et~al.}(2012)\citenamefont {Han},
  \citenamefont {Helton}, \citenamefont {Chu}, \citenamefont {Nocera},
  \citenamefont {Rodriguez-Rivera}, \citenamefont {Broholm},\ and\
  \citenamefont {Lee}}]{Han_2012}%
  \BibitemOpen
  \bibfield  {author} {\bibinfo {author} {\bibfnamefont {T.-H.}\ \bibnamefont
  {Han}}, \bibinfo {author} {\bibfnamefont {J.~S.}\ \bibnamefont {Helton}},
  \bibinfo {author} {\bibfnamefont {S.}~\bibnamefont {Chu}}, \bibinfo {author}
  {\bibfnamefont {D.~G.}\ \bibnamefont {Nocera}}, \bibinfo {author}
  {\bibfnamefont {J.~A.}\ \bibnamefont {Rodriguez-Rivera}}, \bibinfo {author}
  {\bibfnamefont {C.}~\bibnamefont {Broholm}}, \ and\ \bibinfo {author}
  {\bibfnamefont {Y.~S.}\ \bibnamefont {Lee}},\ }\href
  {https://www.nature.com/articles/nature11659} {\bibfield  {journal} {\bibinfo
   {journal} {Nature}\ }\textbf {\bibinfo {volume} {492}},\ \bibinfo {pages}
  {406} (\bibinfo {year} {2012})}\BibitemShut {NoStop}%
\bibitem [{\citenamefont {Savary}\ and\ \citenamefont
  {Balents}(2016)}]{Savary_2016}%
  \BibitemOpen
  \bibfield  {author} {\bibinfo {author} {\bibfnamefont {L.}~\bibnamefont
  {Savary}}\ and\ \bibinfo {author} {\bibfnamefont {L.}~\bibnamefont
  {Balents}},\ }\href
  {http://iopscience.iop.org/article/10.1088/0034-4885/80/1/016502} {\bibfield
  {journal} {\bibinfo  {journal} {Rep. Prog. Phys.}\ }\textbf {\bibinfo
  {volume} {80}},\ \bibinfo {pages} {016502} (\bibinfo {year}
  {2016})}\BibitemShut {NoStop}%
\bibitem [{\citenamefont {Jackeli}\ and\ \citenamefont
  {Khaliullin}(2009)}]{Jackeli_2009}%
  \BibitemOpen
  \bibfield  {author} {\bibinfo {author} {\bibfnamefont {G.}~\bibnamefont
  {Jackeli}}\ and\ \bibinfo {author} {\bibfnamefont {G.}~\bibnamefont
  {Khaliullin}},\ }\href {\doibase 10.1103/PhysRevLett.102.017205} {\bibfield
  {journal} {\bibinfo  {journal} {Phys. Rev. Lett.}\ }\textbf {\bibinfo
  {volume} {102}},\ \bibinfo {pages} {017205} (\bibinfo {year}
  {2009})}\BibitemShut {NoStop}%
\bibitem [{\citenamefont {Banerjee}\ \emph {et~al.}(2017)\citenamefont
  {Banerjee}, \citenamefont {Yan}, \citenamefont {Knolle}, \citenamefont
  {Bridges}, \citenamefont {Stone}, \citenamefont {Lumsden}, \citenamefont
  {Mandrus}, \citenamefont {Tennant}, \citenamefont {Moessner},\ and\
  \citenamefont {Nagler}}]{Banerjee_2016}%
  \BibitemOpen
  \bibfield  {author} {\bibinfo {author} {\bibfnamefont {A.}~\bibnamefont
  {Banerjee}}, \bibinfo {author} {\bibfnamefont {J.}~\bibnamefont {Yan}},
  \bibinfo {author} {\bibfnamefont {J.}~\bibnamefont {Knolle}}, \bibinfo
  {author} {\bibfnamefont {C.~A.}\ \bibnamefont {Bridges}}, \bibinfo {author}
  {\bibfnamefont {M.~B.}\ \bibnamefont {Stone}}, \bibinfo {author}
  {\bibfnamefont {M.~D.}\ \bibnamefont {Lumsden}}, \bibinfo {author}
  {\bibfnamefont {D.~G.}\ \bibnamefont {Mandrus}}, \bibinfo {author}
  {\bibfnamefont {D.~A.}\ \bibnamefont {Tennant}}, \bibinfo {author}
  {\bibfnamefont {R.}~\bibnamefont {Moessner}}, \ and\ \bibinfo {author}
  {\bibfnamefont {S.~E.}\ \bibnamefont {Nagler}},\ }\href {\doibase
  10.1126/science.aah6015} {\bibfield  {journal} {\bibinfo  {journal}
  {Science}\ }\textbf {\bibinfo {volume} {356}},\ \bibinfo {pages} {1055}
  (\bibinfo {year} {2017})}\BibitemShut {NoStop}%
\bibitem [{\citenamefont {Chisnell}\ \emph {et~al.}(2015)\citenamefont
  {Chisnell}, \citenamefont {Helton}, \citenamefont {Freedman}, \citenamefont
  {Singh}, \citenamefont {Bewley}, \citenamefont {Nocera},\ and\ \citenamefont
  {Lee}}]{Chisnell_2015}%
  \BibitemOpen
  \bibfield  {author} {\bibinfo {author} {\bibfnamefont {R.}~\bibnamefont
  {Chisnell}}, \bibinfo {author} {\bibfnamefont {J.~S.}\ \bibnamefont
  {Helton}}, \bibinfo {author} {\bibfnamefont {D.~E.}\ \bibnamefont
  {Freedman}}, \bibinfo {author} {\bibfnamefont {D.~K.}\ \bibnamefont {Singh}},
  \bibinfo {author} {\bibfnamefont {R.~I.}\ \bibnamefont {Bewley}}, \bibinfo
  {author} {\bibfnamefont {D.~G.}\ \bibnamefont {Nocera}}, \ and\ \bibinfo
  {author} {\bibfnamefont {Y.~S.}\ \bibnamefont {Lee}},\ }\href {\doibase
  10.1103/PhysRevLett.115.147201} {\bibfield  {journal} {\bibinfo  {journal}
  {Phys. Rev. Lett.}\ }\textbf {\bibinfo {volume} {115}},\ \bibinfo {pages}
  {147201} (\bibinfo {year} {2015})}\BibitemShut {NoStop}%
\bibitem [{\citenamefont {Hirschberger}\ \emph {et~al.}(2015)\citenamefont
  {Hirschberger}, \citenamefont {Chisnell}, \citenamefont {Lee},\ and\
  \citenamefont {Ong}}]{Hirschberger_2015}%
  \BibitemOpen
  \bibfield  {author} {\bibinfo {author} {\bibfnamefont {M.}~\bibnamefont
  {Hirschberger}}, \bibinfo {author} {\bibfnamefont {R.}~\bibnamefont
  {Chisnell}}, \bibinfo {author} {\bibfnamefont {Y.~S.}\ \bibnamefont {Lee}}, \
  and\ \bibinfo {author} {\bibfnamefont {N.~P.}\ \bibnamefont {Ong}},\ }\href
  {\doibase 10.1103/PhysRevLett.115.106603} {\bibfield  {journal} {\bibinfo
  {journal} {Phys. Rev. Lett.}\ }\textbf {\bibinfo {volume} {115}},\ \bibinfo
  {pages} {106603} (\bibinfo {year} {2015})}\BibitemShut {NoStop}%
\bibitem [{\citenamefont {Chernyshev}\ and\ \citenamefont
  {Maksimov}(2016)}]{Chernyshev_2016}%
  \BibitemOpen
  \bibfield  {author} {\bibinfo {author} {\bibfnamefont {A.~L.}\ \bibnamefont
  {Chernyshev}}\ and\ \bibinfo {author} {\bibfnamefont {P.~A.}\ \bibnamefont
  {Maksimov}},\ }\href {\doibase 10.1103/PhysRevLett.117.187203} {\bibfield
  {journal} {\bibinfo  {journal} {Phys. Rev. Lett.}\ }\textbf {\bibinfo
  {volume} {117}},\ \bibinfo {pages} {187203} (\bibinfo {year}
  {2016})}\BibitemShut {NoStop}%
\bibitem [{\citenamefont {Bramwell}\ and\ \citenamefont
  {Gingras}(2001)}]{Bramwell_2001}%
  \BibitemOpen
  \bibfield  {author} {\bibinfo {author} {\bibfnamefont {S.~T.}\ \bibnamefont
  {Bramwell}}\ and\ \bibinfo {author} {\bibfnamefont {M.~J.}\ \bibnamefont
  {Gingras}},\ }\href {http://science.sciencemag.org/content/294/5546/1495}
  {\bibfield  {journal} {\bibinfo  {journal} {Science}\ }\textbf {\bibinfo
  {volume} {294}},\ \bibinfo {pages} {1495} (\bibinfo {year}
  {2001})}\BibitemShut {NoStop}%
\bibitem [{\citenamefont {Gardner}\ \emph {et~al.}(2010)\citenamefont
  {Gardner}, \citenamefont {Gingras},\ and\ \citenamefont
  {Greedan}}]{Gardner_2010}%
  \BibitemOpen
  \bibfield  {author} {\bibinfo {author} {\bibfnamefont {J.~S.}\ \bibnamefont
  {Gardner}}, \bibinfo {author} {\bibfnamefont {M.~J.}\ \bibnamefont
  {Gingras}}, \ and\ \bibinfo {author} {\bibfnamefont {J.~E.}\ \bibnamefont
  {Greedan}},\ }\href {https://doi.org/10.1103/RevModPhys.82.53} {\bibfield
  {journal} {\bibinfo  {journal} {Rev. Mod. Phys.}\ }\textbf {\bibinfo {volume}
  {82}},\ \bibinfo {pages} {53} (\bibinfo {year} {2010})}\BibitemShut {NoStop}%
\bibitem [{\citenamefont {Fennell}\ \emph {et~al.}(2009)\citenamefont
  {Fennell}, \citenamefont {Deen}, \citenamefont {Wildes}, \citenamefont
  {Schmalzl}, \citenamefont {Prabhakaran}, \citenamefont {Boothroyd},
  \citenamefont {Aldus}, \citenamefont {McMorrow},\ and\ \citenamefont
  {Bramwell}}]{Fennell_2009}%
  \BibitemOpen
  \bibfield  {author} {\bibinfo {author} {\bibfnamefont {T.}~\bibnamefont
  {Fennell}}, \bibinfo {author} {\bibfnamefont {P.}~\bibnamefont {Deen}},
  \bibinfo {author} {\bibfnamefont {A.}~\bibnamefont {Wildes}}, \bibinfo
  {author} {\bibfnamefont {K.}~\bibnamefont {Schmalzl}}, \bibinfo {author}
  {\bibfnamefont {D.}~\bibnamefont {Prabhakaran}}, \bibinfo {author}
  {\bibfnamefont {A.}~\bibnamefont {Boothroyd}}, \bibinfo {author}
  {\bibfnamefont {R.}~\bibnamefont {Aldus}}, \bibinfo {author} {\bibfnamefont
  {D.}~\bibnamefont {McMorrow}}, \ and\ \bibinfo {author} {\bibfnamefont
  {S.}~\bibnamefont {Bramwell}},\ }\href
  {http://science.sciencemag.org/content/326/5951/415} {\bibfield  {journal}
  {\bibinfo  {journal} {Science}\ }\textbf {\bibinfo {volume} {326}},\ \bibinfo
  {pages} {415} (\bibinfo {year} {2009})}\BibitemShut {NoStop}%
\bibitem [{\citenamefont {Ross}\ \emph {et~al.}(2011)\citenamefont {Ross},
  \citenamefont {Savary}, \citenamefont {Gaulin},\ and\ \citenamefont
  {Balents}}]{Ross_2011}%
  \BibitemOpen
  \bibfield  {author} {\bibinfo {author} {\bibfnamefont {K.~A.}\ \bibnamefont
  {Ross}}, \bibinfo {author} {\bibfnamefont {L.}~\bibnamefont {Savary}},
  \bibinfo {author} {\bibfnamefont {B.~D.}\ \bibnamefont {Gaulin}}, \ and\
  \bibinfo {author} {\bibfnamefont {L.}~\bibnamefont {Balents}},\ }\href
  {https://doi.org/10.1103/PhysRevX.1.021002} {\bibfield  {journal} {\bibinfo
  {journal} {Phys. Rev. X}\ }\textbf {\bibinfo {volume} {1}},\ \bibinfo {pages}
  {021002} (\bibinfo {year} {2011})}\BibitemShut {NoStop}%
\bibitem [{\citenamefont {Krimmel}\ \emph {et~al.}(2006)\citenamefont
  {Krimmel}, \citenamefont {M\"ucksch}, \citenamefont {Tsurkan}, \citenamefont
  {Koza}, \citenamefont {Mutka}, \citenamefont {Ritter}, \citenamefont
  {Sheptyakov}, \citenamefont {Horn},\ and\ \citenamefont
  {Loidl}}]{Krimmel_2006}%
  \BibitemOpen
  \bibfield  {author} {\bibinfo {author} {\bibfnamefont {A.}~\bibnamefont
  {Krimmel}}, \bibinfo {author} {\bibfnamefont {M.}~\bibnamefont {M\"ucksch}},
  \bibinfo {author} {\bibfnamefont {V.}~\bibnamefont {Tsurkan}}, \bibinfo
  {author} {\bibfnamefont {M.~M.}\ \bibnamefont {Koza}}, \bibinfo {author}
  {\bibfnamefont {H.}~\bibnamefont {Mutka}}, \bibinfo {author} {\bibfnamefont
  {C.}~\bibnamefont {Ritter}}, \bibinfo {author} {\bibfnamefont {D.~V.}\
  \bibnamefont {Sheptyakov}}, \bibinfo {author} {\bibfnamefont
  {S.}~\bibnamefont {Horn}}, \ and\ \bibinfo {author} {\bibfnamefont
  {A.}~\bibnamefont {Loidl}},\ }\href {\doibase 10.1103/PhysRevB.73.014413}
  {\bibfield  {journal} {\bibinfo  {journal} {Phys. Rev. B}\ }\textbf {\bibinfo
  {volume} {73}},\ \bibinfo {pages} {014413} (\bibinfo {year}
  {2006})}\BibitemShut {NoStop}%
\bibitem [{\citenamefont {Gao}\ \emph {et~al.}(2016)\citenamefont {Gao},
  \citenamefont {Zaharko}, \citenamefont {Tsurkan}, \citenamefont {Su},
  \citenamefont {White}, \citenamefont {Tucker}, \citenamefont {Roessli},
  \citenamefont {Bourdarot}, \citenamefont {Sibille}, \citenamefont
  {Chernyshov}, \citenamefont {Fennell}, \citenamefont {Loidl},\ and\
  \citenamefont {Ruegg}}]{Gao_2016}%
  \BibitemOpen
  \bibfield  {author} {\bibinfo {author} {\bibfnamefont {S.}~\bibnamefont
  {Gao}}, \bibinfo {author} {\bibfnamefont {O.}~\bibnamefont {Zaharko}},
  \bibinfo {author} {\bibfnamefont {V.}~\bibnamefont {Tsurkan}}, \bibinfo
  {author} {\bibfnamefont {Y.}~\bibnamefont {Su}}, \bibinfo {author}
  {\bibfnamefont {J.~S.}\ \bibnamefont {White}}, \bibinfo {author}
  {\bibfnamefont {G.~S.}\ \bibnamefont {Tucker}}, \bibinfo {author}
  {\bibfnamefont {B.}~\bibnamefont {Roessli}}, \bibinfo {author} {\bibfnamefont
  {F.}~\bibnamefont {Bourdarot}}, \bibinfo {author} {\bibfnamefont
  {R.}~\bibnamefont {Sibille}}, \bibinfo {author} {\bibfnamefont
  {D.}~\bibnamefont {Chernyshov}}, \bibinfo {author} {\bibfnamefont
  {T.}~\bibnamefont {Fennell}}, \bibinfo {author} {\bibfnamefont
  {A.}~\bibnamefont {Loidl}}, \ and\ \bibinfo {author} {\bibfnamefont
  {C.}~\bibnamefont {Ruegg}},\ }\href {http://dx.doi.org/10.1038/nphys3914}
  {\bibfield  {journal} {\bibinfo  {journal} {Nat. Phys.}\ }\textbf {\bibinfo
  {volume} {13}},\ \bibinfo {pages} {157–161} (\bibinfo {year}
  {2016})}\BibitemShut {NoStop}%
\bibitem [{\citenamefont {Bergman}\ \emph {et~al.}(2007)\citenamefont
  {Bergman}, \citenamefont {Alicea}, \citenamefont {Gull}, \citenamefont
  {Trebst},\ and\ \citenamefont {Balents}}]{bergman2007order}%
  \BibitemOpen
  \bibfield  {author} {\bibinfo {author} {\bibfnamefont {D.}~\bibnamefont
  {Bergman}}, \bibinfo {author} {\bibfnamefont {J.}~\bibnamefont {Alicea}},
  \bibinfo {author} {\bibfnamefont {E.}~\bibnamefont {Gull}}, \bibinfo {author}
  {\bibfnamefont {S.}~\bibnamefont {Trebst}}, \ and\ \bibinfo {author}
  {\bibfnamefont {L.}~\bibnamefont {Balents}},\ }\href
  {https://www.nature.com/articles/nphys622} {\bibfield  {journal} {\bibinfo
  {journal} {Nat. Phys.}\ }\textbf {\bibinfo {volume} {3}},\ \bibinfo {pages}
  {487} (\bibinfo {year} {2007})}\BibitemShut {NoStop}%
\bibitem [{\citenamefont {Bernier}\ \emph {et~al.}(2008)\citenamefont
  {Bernier}, \citenamefont {Lawler},\ and\ \citenamefont {Kim}}]{Bernier_2008}%
  \BibitemOpen
  \bibfield  {author} {\bibinfo {author} {\bibfnamefont {J.-S.}\ \bibnamefont
  {Bernier}}, \bibinfo {author} {\bibfnamefont {M.~J.}\ \bibnamefont {Lawler}},
  \ and\ \bibinfo {author} {\bibfnamefont {Y.~B.}\ \bibnamefont {Kim}},\ }\href
  {\doibase 10.1103/PhysRevLett.101.047201} {\bibfield  {journal} {\bibinfo
  {journal} {Phys. Rev. Lett.}\ }\textbf {\bibinfo {volume} {101}},\ \bibinfo
  {pages} {047201} (\bibinfo {year} {2008})}\BibitemShut {NoStop}%
\bibitem [{\citenamefont {Fritsch}\ \emph {et~al.}(2004)\citenamefont
  {Fritsch}, \citenamefont {Hemberger}, \citenamefont {B\"uttgen},
  \citenamefont {Scheidt}, \citenamefont {Krug~von Nidda}, \citenamefont
  {Loidl},\ and\ \citenamefont {Tsurkan}}]{Fritsch_2004}%
  \BibitemOpen
  \bibfield  {author} {\bibinfo {author} {\bibfnamefont {V.}~\bibnamefont
  {Fritsch}}, \bibinfo {author} {\bibfnamefont {J.}~\bibnamefont {Hemberger}},
  \bibinfo {author} {\bibfnamefont {N.}~\bibnamefont {B\"uttgen}}, \bibinfo
  {author} {\bibfnamefont {E.-W.}\ \bibnamefont {Scheidt}}, \bibinfo {author}
  {\bibfnamefont {H.-A.}\ \bibnamefont {Krug~von Nidda}}, \bibinfo {author}
  {\bibfnamefont {A.}~\bibnamefont {Loidl}}, \ and\ \bibinfo {author}
  {\bibfnamefont {V.}~\bibnamefont {Tsurkan}},\ }\href {\doibase
  10.1103/PhysRevLett.92.116401} {\bibfield  {journal} {\bibinfo  {journal}
  {Phys. Rev. Lett.}\ }\textbf {\bibinfo {volume} {92}},\ \bibinfo {pages}
  {116401} (\bibinfo {year} {2004})}\BibitemShut {NoStop}%
\bibitem [{\citenamefont {Krimmel}\ \emph {et~al.}(2005)\citenamefont
  {Krimmel}, \citenamefont {M\"ucksch}, \citenamefont {Tsurkan}, \citenamefont
  {Koza}, \citenamefont {Mutka},\ and\ \citenamefont {Loidl}}]{Krimmel_2005}%
  \BibitemOpen
  \bibfield  {author} {\bibinfo {author} {\bibfnamefont {A.}~\bibnamefont
  {Krimmel}}, \bibinfo {author} {\bibfnamefont {M.}~\bibnamefont {M\"ucksch}},
  \bibinfo {author} {\bibfnamefont {V.}~\bibnamefont {Tsurkan}}, \bibinfo
  {author} {\bibfnamefont {M.~M.}\ \bibnamefont {Koza}}, \bibinfo {author}
  {\bibfnamefont {H.}~\bibnamefont {Mutka}}, \ and\ \bibinfo {author}
  {\bibfnamefont {A.}~\bibnamefont {Loidl}},\ }\href {\doibase
  10.1103/PhysRevLett.94.237402} {\bibfield  {journal} {\bibinfo  {journal}
  {Phys. Rev. Lett.}\ }\textbf {\bibinfo {volume} {94}},\ \bibinfo {pages}
  {237402} (\bibinfo {year} {2005})}\BibitemShut {NoStop}%
\bibitem [{\citenamefont {Laurita}\ \emph {et~al.}(2015)\citenamefont
  {Laurita}, \citenamefont {Deisenhofer}, \citenamefont {Pan}, \citenamefont
  {Morris}, \citenamefont {Schmidt}, \citenamefont {Johnsson}, \citenamefont
  {Tsurkan}, \citenamefont {Loidl},\ and\ \citenamefont
  {Armitage}}]{Laurita_2015}%
  \BibitemOpen
  \bibfield  {author} {\bibinfo {author} {\bibfnamefont {N.~J.}\ \bibnamefont
  {Laurita}}, \bibinfo {author} {\bibfnamefont {J.}~\bibnamefont
  {Deisenhofer}}, \bibinfo {author} {\bibfnamefont {L.}~\bibnamefont {Pan}},
  \bibinfo {author} {\bibfnamefont {C.~M.}\ \bibnamefont {Morris}}, \bibinfo
  {author} {\bibfnamefont {M.}~\bibnamefont {Schmidt}}, \bibinfo {author}
  {\bibfnamefont {M.}~\bibnamefont {Johnsson}}, \bibinfo {author}
  {\bibfnamefont {V.}~\bibnamefont {Tsurkan}}, \bibinfo {author} {\bibfnamefont
  {A.}~\bibnamefont {Loidl}}, \ and\ \bibinfo {author} {\bibfnamefont {N.~P.}\
  \bibnamefont {Armitage}},\ }\href {\doibase 10.1103/PhysRevLett.114.207201}
  {\bibfield  {journal} {\bibinfo  {journal} {Phys. Rev. Lett.}\ }\textbf
  {\bibinfo {volume} {114}},\ \bibinfo {pages} {207201} (\bibinfo {year}
  {2015})}\BibitemShut {NoStop}%
\bibitem [{\citenamefont {Mittelst\"adt}\ \emph {et~al.}(2015)\citenamefont
  {Mittelst\"adt}, \citenamefont {Schmidt}, \citenamefont {Wang}, \citenamefont
  {Mayr}, \citenamefont {Tsurkan}, \citenamefont {Lunkenheimer}, \citenamefont
  {Ish}, \citenamefont {Balents}, \citenamefont {Deisenhofer},\ and\
  \citenamefont {Loidl}}]{Mittelstadt_2015}%
  \BibitemOpen
  \bibfield  {author} {\bibinfo {author} {\bibfnamefont {L.}~\bibnamefont
  {Mittelst\"adt}}, \bibinfo {author} {\bibfnamefont {M.}~\bibnamefont
  {Schmidt}}, \bibinfo {author} {\bibfnamefont {Z.}~\bibnamefont {Wang}},
  \bibinfo {author} {\bibfnamefont {F.}~\bibnamefont {Mayr}}, \bibinfo {author}
  {\bibfnamefont {V.}~\bibnamefont {Tsurkan}}, \bibinfo {author} {\bibfnamefont
  {P.}~\bibnamefont {Lunkenheimer}}, \bibinfo {author} {\bibfnamefont
  {D.}~\bibnamefont {Ish}}, \bibinfo {author} {\bibfnamefont {L.}~\bibnamefont
  {Balents}}, \bibinfo {author} {\bibfnamefont {J.}~\bibnamefont
  {Deisenhofer}}, \ and\ \bibinfo {author} {\bibfnamefont {A.}~\bibnamefont
  {Loidl}},\ }\href {\doibase 10.1103/PhysRevB.91.125112} {\bibfield  {journal}
  {\bibinfo  {journal} {Phys. Rev. B}\ }\textbf {\bibinfo {volume} {91}},\
  \bibinfo {pages} {125112} (\bibinfo {year} {2015})}\BibitemShut {NoStop}%
\bibitem [{\citenamefont {Plumb}\ \emph {et~al.}(2016)\citenamefont {Plumb},
  \citenamefont {Morey}, \citenamefont {Rodriguez-Rivera}, \citenamefont {Wu},
  \citenamefont {Podlesnyak}, \citenamefont {McQueen},\ and\ \citenamefont
  {Broholm}}]{Plumb_2016}%
  \BibitemOpen
  \bibfield  {author} {\bibinfo {author} {\bibfnamefont {K.~W.}\ \bibnamefont
  {Plumb}}, \bibinfo {author} {\bibfnamefont {J.~R.}\ \bibnamefont {Morey}},
  \bibinfo {author} {\bibfnamefont {J.~A.}\ \bibnamefont {Rodriguez-Rivera}},
  \bibinfo {author} {\bibfnamefont {H.}~\bibnamefont {Wu}}, \bibinfo {author}
  {\bibfnamefont {A.~A.}\ \bibnamefont {Podlesnyak}}, \bibinfo {author}
  {\bibfnamefont {T.~M.}\ \bibnamefont {McQueen}}, \ and\ \bibinfo {author}
  {\bibfnamefont {C.~L.}\ \bibnamefont {Broholm}},\ }\href {\doibase
  10.1103/PhysRevX.6.041055} {\bibfield  {journal} {\bibinfo  {journal} {Phys.
  Rev. X}\ }\textbf {\bibinfo {volume} {6}},\ \bibinfo {pages} {041055}
  (\bibinfo {year} {2016})}\BibitemShut {NoStop}%
\bibitem [{\citenamefont {Biffin}\ \emph {et~al.}(2017)\citenamefont {Biffin},
  \citenamefont {R\"uegg}, \citenamefont {Embs}, \citenamefont {Guidi},
  \citenamefont {Cheptiakov}, \citenamefont {Loidl}, \citenamefont {Tsurkan},\
  and\ \citenamefont {Coldea}}]{Biffin_2017}%
  \BibitemOpen
  \bibfield  {author} {\bibinfo {author} {\bibfnamefont {A.}~\bibnamefont
  {Biffin}}, \bibinfo {author} {\bibfnamefont {C.}~\bibnamefont {R\"uegg}},
  \bibinfo {author} {\bibfnamefont {J.}~\bibnamefont {Embs}}, \bibinfo {author}
  {\bibfnamefont {T.}~\bibnamefont {Guidi}}, \bibinfo {author} {\bibfnamefont
  {D.}~\bibnamefont {Cheptiakov}}, \bibinfo {author} {\bibfnamefont
  {A.}~\bibnamefont {Loidl}}, \bibinfo {author} {\bibfnamefont
  {V.}~\bibnamefont {Tsurkan}}, \ and\ \bibinfo {author} {\bibfnamefont
  {R.}~\bibnamefont {Coldea}},\ }\href {\doibase
  10.1103/PhysRevLett.118.067205} {\bibfield  {journal} {\bibinfo  {journal}
  {Phys. Rev. Lett.}\ }\textbf {\bibinfo {volume} {118}},\ \bibinfo {pages}
  {067205} (\bibinfo {year} {2017})}\BibitemShut {NoStop}%
\bibitem [{\citenamefont {Chen}\ \emph
  {et~al.}(2009{\natexlab{a}})\citenamefont {Chen}, \citenamefont {Balents},\
  and\ \citenamefont {Schnyder}}]{Chen_2009a}%
  \BibitemOpen
  \bibfield  {author} {\bibinfo {author} {\bibfnamefont {G.}~\bibnamefont
  {Chen}}, \bibinfo {author} {\bibfnamefont {L.}~\bibnamefont {Balents}}, \
  and\ \bibinfo {author} {\bibfnamefont {A.~P.}\ \bibnamefont {Schnyder}},\
  }\href {\doibase 10.1103/PhysRevLett.102.096406} {\bibfield  {journal}
  {\bibinfo  {journal} {Phys. Rev. Lett.}\ }\textbf {\bibinfo {volume} {102}},\
  \bibinfo {pages} {096406} (\bibinfo {year} {2009}{\natexlab{a}})}\BibitemShut
  {NoStop}%
\bibitem [{\citenamefont {Chen}\ \emph
  {et~al.}(2009{\natexlab{b}})\citenamefont {Chen}, \citenamefont {Schnyder},\
  and\ \citenamefont {Balents}}]{Chen_2009b}%
  \BibitemOpen
  \bibfield  {author} {\bibinfo {author} {\bibfnamefont {G.}~\bibnamefont
  {Chen}}, \bibinfo {author} {\bibfnamefont {A.~P.}\ \bibnamefont {Schnyder}},
  \ and\ \bibinfo {author} {\bibfnamefont {L.}~\bibnamefont {Balents}},\ }\href
  {\doibase 10.1103/PhysRevB.80.224409} {\bibfield  {journal} {\bibinfo
  {journal} {Phys. Rev. B}\ }\textbf {\bibinfo {volume} {80}},\ \bibinfo
  {pages} {224409} (\bibinfo {year} {2009}{\natexlab{b}})}\BibitemShut
  {NoStop}%
\bibitem [{\citenamefont {Suzuki}\ \emph {et~al.}(2007)\citenamefont {Suzuki},
  \citenamefont {Nagai}, \citenamefont {Nohara},\ and\ \citenamefont
  {Takagi}}]{Suzuki_2007}%
  \BibitemOpen
  \bibfield  {author} {\bibinfo {author} {\bibfnamefont {T.}~\bibnamefont
  {Suzuki}}, \bibinfo {author} {\bibfnamefont {H.}~\bibnamefont {Nagai}},
  \bibinfo {author} {\bibfnamefont {M.}~\bibnamefont {Nohara}}, \ and\ \bibinfo
  {author} {\bibfnamefont {H.}~\bibnamefont {Takagi}},\ }\href
  {http://stacks.iop.org/0953-8984/19/i=14/a=145265} {\bibfield  {journal}
  {\bibinfo  {journal} {J. Phys. Condens. Matter}\ }\textbf {\bibinfo {volume}
  {19}},\ \bibinfo {pages} {145265} (\bibinfo {year} {2007})}\BibitemShut
  {NoStop}%
\bibitem [{\citenamefont {MacDougall}\ \emph {et~al.}(2011)\citenamefont
  {MacDougall}, \citenamefont {Gout}, \citenamefont {Zarestky}, \citenamefont
  {Ehlers}, \citenamefont {Podlesnyak}, \citenamefont {McGuire}, \citenamefont
  {Mandrus},\ and\ \citenamefont {Nagler}}]{MacDougall20092011}%
  \BibitemOpen
  \bibfield  {author} {\bibinfo {author} {\bibfnamefont {G.~J.}\ \bibnamefont
  {MacDougall}}, \bibinfo {author} {\bibfnamefont {D.}~\bibnamefont {Gout}},
  \bibinfo {author} {\bibfnamefont {J.~L.}\ \bibnamefont {Zarestky}}, \bibinfo
  {author} {\bibfnamefont {G.}~\bibnamefont {Ehlers}}, \bibinfo {author}
  {\bibfnamefont {A.}~\bibnamefont {Podlesnyak}}, \bibinfo {author}
  {\bibfnamefont {M.~A.}\ \bibnamefont {McGuire}}, \bibinfo {author}
  {\bibfnamefont {D.}~\bibnamefont {Mandrus}}, \ and\ \bibinfo {author}
  {\bibfnamefont {S.~E.}\ \bibnamefont {Nagler}},\ }\href {\doibase
  10.1073/pnas.1107861108} {\bibfield  {journal} {\bibinfo  {journal} {Proc.
  Natl. Acad. Sci. USA}\ }\textbf {\bibinfo {volume} {108}},\ \bibinfo {pages}
  {15693} (\bibinfo {year} {2011})}\BibitemShut {NoStop}%
\bibitem [{\citenamefont {Zaharko}\ \emph {et~al.}(2014)\citenamefont
  {Zaharko}, \citenamefont {T\'oth}, \citenamefont {Sendetskyi}, \citenamefont
  {Cervellino}, \citenamefont {Wolter-Giraud}, \citenamefont {Dey},
  \citenamefont {Maljuk},\ and\ \citenamefont {Tsurkan}}]{Zaharko_2014}%
  \BibitemOpen
  \bibfield  {author} {\bibinfo {author} {\bibfnamefont {O.}~\bibnamefont
  {Zaharko}}, \bibinfo {author} {\bibfnamefont {S.}~\bibnamefont {T\'oth}},
  \bibinfo {author} {\bibfnamefont {O.}~\bibnamefont {Sendetskyi}}, \bibinfo
  {author} {\bibfnamefont {A.}~\bibnamefont {Cervellino}}, \bibinfo {author}
  {\bibfnamefont {A.}~\bibnamefont {Wolter-Giraud}}, \bibinfo {author}
  {\bibfnamefont {T.}~\bibnamefont {Dey}}, \bibinfo {author} {\bibfnamefont
  {A.}~\bibnamefont {Maljuk}}, \ and\ \bibinfo {author} {\bibfnamefont
  {V.}~\bibnamefont {Tsurkan}},\ }\href {\doibase 10.1103/PhysRevB.90.134416}
  {\bibfield  {journal} {\bibinfo  {journal} {Phys. Rev. B}\ }\textbf {\bibinfo
  {volume} {90}},\ \bibinfo {pages} {134416} (\bibinfo {year}
  {2014})}\BibitemShut {NoStop}%
\bibitem [{\citenamefont {MacDougall}\ \emph {et~al.}(2016)\citenamefont
  {MacDougall}, \citenamefont {Aczel}, \citenamefont {Su}, \citenamefont
  {Schweika}, \citenamefont {Faulhaber}, \citenamefont {Schneidewind},
  \citenamefont {Christianson}, \citenamefont {Zarestky}, \citenamefont {Zhou},
  \citenamefont {Mandrus},\ and\ \citenamefont {Nagler}}]{MacDougall_2016}%
  \BibitemOpen
  \bibfield  {author} {\bibinfo {author} {\bibfnamefont {G.~J.}\ \bibnamefont
  {MacDougall}}, \bibinfo {author} {\bibfnamefont {A.~A.}\ \bibnamefont
  {Aczel}}, \bibinfo {author} {\bibfnamefont {Y.}~\bibnamefont {Su}}, \bibinfo
  {author} {\bibfnamefont {W.}~\bibnamefont {Schweika}}, \bibinfo {author}
  {\bibfnamefont {E.}~\bibnamefont {Faulhaber}}, \bibinfo {author}
  {\bibfnamefont {A.}~\bibnamefont {Schneidewind}}, \bibinfo {author}
  {\bibfnamefont {A.~D.}\ \bibnamefont {Christianson}}, \bibinfo {author}
  {\bibfnamefont {J.~L.}\ \bibnamefont {Zarestky}}, \bibinfo {author}
  {\bibfnamefont {H.~D.}\ \bibnamefont {Zhou}}, \bibinfo {author}
  {\bibfnamefont {D.}~\bibnamefont {Mandrus}}, \ and\ \bibinfo {author}
  {\bibfnamefont {S.~E.}\ \bibnamefont {Nagler}},\ }\href {\doibase
  10.1103/PhysRevB.94.184422} {\bibfield  {journal} {\bibinfo  {journal} {Phys.
  Rev. B}\ }\textbf {\bibinfo {volume} {94}},\ \bibinfo {pages} {184422}
  (\bibinfo {year} {2016})}\BibitemShut {NoStop}%
\bibitem [{\citenamefont {Wang}\ \emph {et~al.}(2015)\citenamefont {Wang},
  \citenamefont {Nahum},\ and\ \citenamefont {Senthil}}]{Senthil_2015}%
  \BibitemOpen
  \bibfield  {author} {\bibinfo {author} {\bibfnamefont {C.}~\bibnamefont
  {Wang}}, \bibinfo {author} {\bibfnamefont {A.}~\bibnamefont {Nahum}}, \ and\
  \bibinfo {author} {\bibfnamefont {T.}~\bibnamefont {Senthil}},\ }\href
  {\doibase 10.1103/PhysRevB.91.195131} {\bibfield  {journal} {\bibinfo
  {journal} {Phys. Rev. B}\ }\textbf {\bibinfo {volume} {91}},\ \bibinfo
  {pages} {195131} (\bibinfo {year} {2015})}\BibitemShut {NoStop}%
\bibitem [{\citenamefont {Haldane}(1983)}]{Haldane_1983}%
  \BibitemOpen
  \bibfield  {author} {\bibinfo {author} {\bibfnamefont {F.~D.~M.}\
  \bibnamefont {Haldane}},\ }\href {\doibase 10.1103/PhysRevLett.50.1153}
  {\bibfield  {journal} {\bibinfo  {journal} {Phys. Rev. Lett.}\ }\textbf
  {\bibinfo {volume} {50}},\ \bibinfo {pages} {1153} (\bibinfo {year}
  {1983})}\BibitemShut {NoStop}%
\bibitem [{\citenamefont {Affleck}\ \emph {et~al.}(1987)\citenamefont
  {Affleck}, \citenamefont {Kennedy}, \citenamefont {Lieb},\ and\ \citenamefont
  {Tasaki}}]{AKLT}%
  \BibitemOpen
  \bibfield  {author} {\bibinfo {author} {\bibfnamefont {I.}~\bibnamefont
  {Affleck}}, \bibinfo {author} {\bibfnamefont {T.}~\bibnamefont {Kennedy}},
  \bibinfo {author} {\bibfnamefont {E.~H.}\ \bibnamefont {Lieb}}, \ and\
  \bibinfo {author} {\bibfnamefont {H.}~\bibnamefont {Tasaki}},\ }\href
  {\doibase 10.1103/PhysRevLett.59.799} {\bibfield  {journal} {\bibinfo
  {journal} {Phys. Rev. Lett.}\ }\textbf {\bibinfo {volume} {59}},\ \bibinfo
  {pages} {799} (\bibinfo {year} {1987})}\BibitemShut {NoStop}%
\bibitem [{\citenamefont {Chamorro}\ and\ \citenamefont
  {McQueen}(2017)}]{chamorro2017frustrated}%
  \BibitemOpen
  \bibfield  {author} {\bibinfo {author} {\bibfnamefont {J.~R.}\ \bibnamefont
  {Chamorro}}\ and\ \bibinfo {author} {\bibfnamefont {T.~M.}\ \bibnamefont
  {McQueen}},\ }\href {https://arxiv.org/abs/1701.06674} {\bibfield  {journal}
  {\bibinfo  {journal} {arXiv preprint arXiv:1701.06674}\ } (\bibinfo {year}
  {2017})}\BibitemShut {NoStop}%
\bibitem [{\citenamefont {Chen}(2017)}]{chen2017quantum}%
  \BibitemOpen
  \bibfield  {author} {\bibinfo {author} {\bibfnamefont {G.}~\bibnamefont
  {Chen}},\ }\href {https://arxiv.org/abs/1701.05634} {\bibfield  {journal}
  {\bibinfo  {journal} {arXiv preprint arXiv:1701.05634}\ } (\bibinfo {year}
  {2017})}\BibitemShut {NoStop}%
\bibitem [{\citenamefont {Rodriguez-Carvajal}(1993)}]{Carvajal_1993}%
  \BibitemOpen
  \bibfield  {author} {\bibinfo {author} {\bibfnamefont {J.}~\bibnamefont
  {Rodriguez-Carvajal}},\ }\href {\doibase
  http://dx.doi.org/10.1016/0921-4526(93)90108-I} {\bibfield  {journal}
  {\bibinfo  {journal} {Physica B}\ }\textbf {\bibinfo {volume} {192}},\
  \bibinfo {pages} {55 } (\bibinfo {year} {1993})}\BibitemShut {NoStop}%
\bibitem [{\citenamefont {Bain}\ and\ \citenamefont {Berry}(2008)}]{Bain_2008}%
  \BibitemOpen
  \bibfield  {author} {\bibinfo {author} {\bibfnamefont {G.~A.}\ \bibnamefont
  {Bain}}\ and\ \bibinfo {author} {\bibfnamefont {J.~F.}\ \bibnamefont
  {Berry}},\ }\href {\doibase 10.1021/ed085p532} {\bibfield  {journal}
  {\bibinfo  {journal} {Journal of Chemical Education}\ }\textbf {\bibinfo
  {volume} {85}},\ \bibinfo {pages} {532} (\bibinfo {year} {2008})}\BibitemShut
  {NoStop}%
\bibitem [{\citenamefont {Garlea}\ \emph {et~al.}(2010)\citenamefont {Garlea},
  \citenamefont {Chakoumakos}, \citenamefont {Moore}, \citenamefont {Taylor},
  \citenamefont {Chae}, \citenamefont {Maples}, \citenamefont {Riedel},
  \citenamefont {Lynn},\ and\ \citenamefont {Selby}}]{Garlea_2010}%
  \BibitemOpen
  \bibfield  {author} {\bibinfo {author} {\bibfnamefont {V.~O.}\ \bibnamefont
  {Garlea}}, \bibinfo {author} {\bibfnamefont {B.~C.}\ \bibnamefont
  {Chakoumakos}}, \bibinfo {author} {\bibfnamefont {S.~A.}\ \bibnamefont
  {Moore}}, \bibinfo {author} {\bibfnamefont {G.~B.}\ \bibnamefont {Taylor}},
  \bibinfo {author} {\bibfnamefont {T.}~\bibnamefont {Chae}}, \bibinfo {author}
  {\bibfnamefont {R.~G.}\ \bibnamefont {Maples}}, \bibinfo {author}
  {\bibfnamefont {R.~A.}\ \bibnamefont {Riedel}}, \bibinfo {author}
  {\bibfnamefont {G.~W.}\ \bibnamefont {Lynn}}, \ and\ \bibinfo {author}
  {\bibfnamefont {D.~L.}\ \bibnamefont {Selby}},\ }\href {\doibase
  10.1007/s00339-010-5603-6} {\bibfield  {journal} {\bibinfo  {journal}
  {Applied Physics A}\ }\textbf {\bibinfo {volume} {99}},\ \bibinfo {pages}
  {531} (\bibinfo {year} {2010})}\BibitemShut {NoStop}%
\bibitem [{\citenamefont {Granroth}\ \emph {et~al.}(2010)\citenamefont
  {Granroth}, \citenamefont {Kolesnikov}, \citenamefont {Sherline},
  \citenamefont {Clancy}, \citenamefont {Ross}, \citenamefont {Ruff},
  \citenamefont {Gaulin},\ and\ \citenamefont {Nagler}}]{Granroth_2010}%
  \BibitemOpen
  \bibfield  {author} {\bibinfo {author} {\bibfnamefont {G.~E.}\ \bibnamefont
  {Granroth}}, \bibinfo {author} {\bibfnamefont {A.~I.}\ \bibnamefont
  {Kolesnikov}}, \bibinfo {author} {\bibfnamefont {T.~E.}\ \bibnamefont
  {Sherline}}, \bibinfo {author} {\bibfnamefont {J.~P.}\ \bibnamefont
  {Clancy}}, \bibinfo {author} {\bibfnamefont {K.~A.}\ \bibnamefont {Ross}},
  \bibinfo {author} {\bibfnamefont {J.~P.~C.}\ \bibnamefont {Ruff}}, \bibinfo
  {author} {\bibfnamefont {B.~D.}\ \bibnamefont {Gaulin}}, \ and\ \bibinfo
  {author} {\bibfnamefont {S.~E.}\ \bibnamefont {Nagler}},\ }\href
  {http://stacks.iop.org/1742-6596/251/i=1/a=012058} {\bibfield  {journal}
  {\bibinfo  {journal} {J. Phys.: Conf. Series}\ }\textbf {\bibinfo {volume}
  {251}},\ \bibinfo {pages} {012058} (\bibinfo {year} {2010})}\BibitemShut
  {NoStop}%
\bibitem [{\citenamefont {Stone}\ \emph {et~al.}(2014)\citenamefont {Stone},
  \citenamefont {Niedziela}, \citenamefont {Abernathy}, \citenamefont
  {DeBeer-Schmitt}, \citenamefont {Ehlers}, \citenamefont {Garlea},
  \citenamefont {Granroth}, \citenamefont {Graves-Brook}, \citenamefont
  {Kolesnikov}, \citenamefont {Podlesnyak},\ and\ \citenamefont
  {Winn}}]{Stone_2014}%
  \BibitemOpen
  \bibfield  {author} {\bibinfo {author} {\bibfnamefont {M.~B.}\ \bibnamefont
  {Stone}}, \bibinfo {author} {\bibfnamefont {J.~L.}\ \bibnamefont
  {Niedziela}}, \bibinfo {author} {\bibfnamefont {D.~L.}\ \bibnamefont
  {Abernathy}}, \bibinfo {author} {\bibfnamefont {L.}~\bibnamefont
  {DeBeer-Schmitt}}, \bibinfo {author} {\bibfnamefont {G.}~\bibnamefont
  {Ehlers}}, \bibinfo {author} {\bibfnamefont {O.}~\bibnamefont {Garlea}},
  \bibinfo {author} {\bibfnamefont {G.~E.}\ \bibnamefont {Granroth}}, \bibinfo
  {author} {\bibfnamefont {M.}~\bibnamefont {Graves-Brook}}, \bibinfo {author}
  {\bibfnamefont {A.~I.}\ \bibnamefont {Kolesnikov}}, \bibinfo {author}
  {\bibfnamefont {A.}~\bibnamefont {Podlesnyak}}, \ and\ \bibinfo {author}
  {\bibfnamefont {B.}~\bibnamefont {Winn}},\ }\href {\doibase
  10.1063/1.4870050} {\bibfield  {journal} {\bibinfo  {journal} {Rev. Sci.
  Instrum.}\ }\textbf {\bibinfo {volume} {85}},\ \bibinfo {pages} {045113}
  (\bibinfo {year} {2014})}\BibitemShut {NoStop}%
\bibitem [{\citenamefont {{Petit, S}}(2011)}]{Petit_2011}%
  \BibitemOpen
  \bibfield  {author} {\bibinfo {author} {\bibnamefont {{Petit, S}}},\ }\href
  {\doibase 10.1051/sfn/201112006} {\bibfield  {journal} {\bibinfo  {journal}
  {JDN}\ }\textbf {\bibinfo {volume} {12}},\ \bibinfo {pages} {105} (\bibinfo
  {year} {2011})}\BibitemShut {NoStop}%
\bibitem [{\citenamefont {Toth}\ and\ \citenamefont {Lake}(2015)}]{Toth_2015}%
  \BibitemOpen
  \bibfield  {author} {\bibinfo {author} {\bibfnamefont {S.}~\bibnamefont
  {Toth}}\ and\ \bibinfo {author} {\bibfnamefont {B.}~\bibnamefont {Lake}},\
  }\href {http://stacks.iop.org/0953-8984/27/i=16/a=166002} {\bibfield
  {journal} {\bibinfo  {journal} {J. Phys. Condens. Matter}\ }\textbf {\bibinfo
  {volume} {27}},\ \bibinfo {pages} {166002} (\bibinfo {year}
  {2015})}\BibitemShut {NoStop}%
\bibitem [{\citenamefont {Blasse}(1963)}]{Blasse_1963b}%
  \BibitemOpen
  \bibfield  {author} {\bibinfo {author} {\bibfnamefont {G.}~\bibnamefont
  {Blasse}},\ }\href {https://doi.org/10.1016/0022-3697(66)90045-X} {\bibfield
  {journal} {\bibinfo  {journal} {Philips Res. Repts}\ }\textbf {\bibinfo
  {volume} {18}},\ \bibinfo {pages} {383} (\bibinfo {year} {1963})}\BibitemShut
  {NoStop}%
\bibitem [{\citenamefont {Bertaut}\ \emph {et~al.}(1959)\citenamefont
  {Bertaut}, \citenamefont {Forrat},\ and\ \citenamefont
  {Dulac}}]{Bertaut_1959}%
  \BibitemOpen
  \bibfield  {author} {\bibinfo {author} {\bibfnamefont {F.}~\bibnamefont
  {Bertaut}}, \bibinfo {author} {\bibfnamefont {F.}~\bibnamefont {Forrat}}, \
  and\ \bibinfo {author} {\bibfnamefont {J.}~\bibnamefont {Dulac}},\
  }\href@noop {} {\bibfield  {journal} {\bibinfo  {journal} {Comptes Rendus
  Hebdomadaires des Seances de l'Academie des Sciences}\ }\textbf {\bibinfo
  {volume} {249}},\ \bibinfo {pages} {726} (\bibinfo {year}
  {1959})}\BibitemShut {NoStop}%
\bibitem [{\citenamefont {Cascales}\ and\ \citenamefont
  {Rasines}(1984)}]{Cascales_1984}%
  \BibitemOpen
  \bibfield  {author} {\bibinfo {author} {\bibfnamefont {C.}~\bibnamefont
  {Cascales}}\ and\ \bibinfo {author} {\bibfnamefont {I.}~\bibnamefont
  {Rasines}},\ }\href {https://doi.org/10.1016/0254-0584(84)90048-8} {\bibfield
   {journal} {\bibinfo  {journal} {Materials Chemistry And Physics}\ }\textbf
  {\bibinfo {volume} {10}},\ \bibinfo {pages} {199} (\bibinfo {year}
  {1984})}\BibitemShut {NoStop}%
\bibitem [{\citenamefont {Brown}(1981)}]{Brown_1981}%
  \BibitemOpen
  \bibfield  {author} {\bibinfo {author} {\bibfnamefont {I.~D.}\ \bibnamefont
  {Brown}},\ }\href@noop {} {\emph {\bibinfo {title} {The bond-valence method:
  an empirical approach to chemical structure and bonding}}}\ (\bibinfo
  {publisher} {Academic Press},\ \bibinfo {address} {New York},\ \bibinfo
  {year} {1981})\ pp.\ \bibinfo {pages} {1--30}\BibitemShut {NoStop}%
\bibitem [{\citenamefont {Blasse}\ and\ \citenamefont
  {Schipper}(1963)}]{Blasse_1963}%
  \BibitemOpen
  \bibfield  {author} {\bibinfo {author} {\bibfnamefont {G.}~\bibnamefont
  {Blasse}}\ and\ \bibinfo {author} {\bibfnamefont {D.~J.}\ \bibnamefont
  {Schipper}},\ }\href {https://doi.org/10.1016/S0375-9601(63)93851-9}
  {\bibfield  {journal} {\bibinfo  {journal} {Phys. Lett.}\ }\textbf {\bibinfo
  {volume} {5}},\ \bibinfo {pages} {300} (\bibinfo {year} {1963})}\BibitemShut
  {NoStop}%
\bibitem [{\citenamefont {Fiorani}\ and\ \citenamefont
  {Viticoli}(1979)}]{Fiorani_1979}%
  \BibitemOpen
  \bibfield  {author} {\bibinfo {author} {\bibfnamefont {D.}~\bibnamefont
  {Fiorani}}\ and\ \bibinfo {author} {\bibfnamefont {S.}~\bibnamefont
  {Viticoli}},\ }\href {https://doi.org/10.1016/0038-1098(79)91046-9}
  {\bibfield  {journal} {\bibinfo  {journal} {Solid State Commun.}\ }\textbf
  {\bibinfo {volume} {29}},\ \bibinfo {pages} {239} (\bibinfo {year}
  {1979})}\BibitemShut {NoStop}%
\bibitem [{\citenamefont {Campbell}\ \emph {et~al.}(2006)\citenamefont
  {Campbell}, \citenamefont {Stokes}, \citenamefont {Tanner},\ and\
  \citenamefont {Hatch}}]{Campbell_2006}%
  \BibitemOpen
  \bibfield  {author} {\bibinfo {author} {\bibfnamefont {B.~J.}\ \bibnamefont
  {Campbell}}, \bibinfo {author} {\bibfnamefont {H.~T.}\ \bibnamefont
  {Stokes}}, \bibinfo {author} {\bibfnamefont {D.~E.}\ \bibnamefont {Tanner}},
  \ and\ \bibinfo {author} {\bibfnamefont {D.~M.}\ \bibnamefont {Hatch}},\
  }\href {\doibase 10.1107/S0021889806014075} {\bibfield  {journal} {\bibinfo
  {journal} {J. Appl. Crystallogr.}\ }\textbf {\bibinfo {volume} {39}},\
  \bibinfo {pages} {607} (\bibinfo {year} {2006})}\BibitemShut {NoStop}%
\bibitem [{\citenamefont {Miller}\ and\ \citenamefont
  {Love}(1967)}]{Miller_1967}%
  \BibitemOpen
  \bibfield  {author} {\bibinfo {author} {\bibfnamefont {S.~C.}\ \bibnamefont
  {Miller}}\ and\ \bibinfo {author} {\bibfnamefont {W.~F.}\ \bibnamefont
  {Love}},\ }\href@noop {} {\emph {\bibinfo {title} {Tables of Irreducible
  Representations of Space Groups and Co-representations of Magnetic Space
  Groups}}}\ (\bibinfo  {publisher} {Pruett Press},\ \bibinfo {address}
  {Boulder, Colorado},\ \bibinfo {year} {1967})\BibitemShut {NoStop}%
\bibitem [{\citenamefont {R\o{}nnow}\ \emph {et~al.}(1999)\citenamefont
  {R\o{}nnow}, \citenamefont {McMorrow},\ and\ \citenamefont
  {Harrison}}]{Ronnow_1999}%
  \BibitemOpen
  \bibfield  {author} {\bibinfo {author} {\bibfnamefont {H.~M.}\ \bibnamefont
  {R\o{}nnow}}, \bibinfo {author} {\bibfnamefont {D.~F.}\ \bibnamefont
  {McMorrow}}, \ and\ \bibinfo {author} {\bibfnamefont {A.}~\bibnamefont
  {Harrison}},\ }\href {\doibase 10.1103/PhysRevLett.82.3152} {\bibfield
  {journal} {\bibinfo  {journal} {Phys. Rev. Lett.}\ }\textbf {\bibinfo
  {volume} {82}},\ \bibinfo {pages} {3152} (\bibinfo {year}
  {1999})}\BibitemShut {NoStop}%
\bibitem [{\citenamefont {Khanolkar}(1961)}]{Khanolkar_1961}%
  \BibitemOpen
  \bibfield  {author} {\bibinfo {author} {\bibfnamefont {D.}~\bibnamefont
  {Khanolkar}},\ }\href@noop {} {\bibfield  {journal} {\bibinfo  {journal}
  {Current Science}\ }\textbf {\bibinfo {volume} {30}},\ \bibinfo {pages} {52}
  (\bibinfo {year} {1961})}\BibitemShut {NoStop}%
\bibitem [{\citenamefont {Ismunandar}\ \emph {et~al.}(1999)\citenamefont
  {Ismunandar}, \citenamefont {Kennedy},\ and\ \citenamefont
  {Hunter}}]{Kennedy_1999}%
  \BibitemOpen
  \bibfield  {author} {\bibinfo {author} {\bibnamefont {Ismunandar}}, \bibinfo
  {author} {\bibfnamefont {B.~J.}\ \bibnamefont {Kennedy}}, \ and\ \bibinfo
  {author} {\bibfnamefont {B.~A.}\ \bibnamefont {Hunter}},\ }\href
  {https://doi.org/10.1016/S0025-5408(98)00201-3} {\bibfield  {journal}
  {\bibinfo  {journal} {Mat. Res. Bull.}\ }\textbf {\bibinfo {volume} {34}},\
  \bibinfo {pages} {135} (\bibinfo {year} {1999})}\BibitemShut {NoStop}%
\bibitem [{\citenamefont {Dollase}\ and\ \citenamefont
  {O'Neill}(1997)}]{Dollase_1997}%
  \BibitemOpen
  \bibfield  {author} {\bibinfo {author} {\bibfnamefont {W.}~\bibnamefont
  {Dollase}}\ and\ \bibinfo {author} {\bibfnamefont {H.~S.~C.}\ \bibnamefont
  {O'Neill}},\ }\href {https://doi.org/10.1107/S0108270197000486} {\bibfield
  {journal} {\bibinfo  {journal} {Acta Crystallogr. Sec. C}\ }\textbf {\bibinfo
  {volume} {53}},\ \bibinfo {pages} {657} (\bibinfo {year} {1997})}\BibitemShut
  {NoStop}%
\bibitem [{\citenamefont {Endoh}\ \emph {et~al.}(1999)\citenamefont {Endoh},
  \citenamefont {Fujishima}, \citenamefont {Atake}, \citenamefont {Matsumoto},
  \citenamefont {Hayashi},\ and\ \citenamefont {Nagata}}]{Endoh_1999}%
  \BibitemOpen
  \bibfield  {author} {\bibinfo {author} {\bibfnamefont {R.}~\bibnamefont
  {Endoh}}, \bibinfo {author} {\bibfnamefont {O.}~\bibnamefont {Fujishima}},
  \bibinfo {author} {\bibfnamefont {T.}~\bibnamefont {Atake}}, \bibinfo
  {author} {\bibfnamefont {N.}~\bibnamefont {Matsumoto}}, \bibinfo {author}
  {\bibfnamefont {M.}~\bibnamefont {Hayashi}}, \ and\ \bibinfo {author}
  {\bibfnamefont {S.}~\bibnamefont {Nagata}},\ }\href {\doibase
  http://dx.doi.org/10.1016/S0022-3697(98)00310-2} {\bibfield  {journal}
  {\bibinfo  {journal} {J. Phys. Chem. Solids}\ }\textbf {\bibinfo {volume}
  {60}},\ \bibinfo {pages} {457 } (\bibinfo {year} {1999})}\BibitemShut
  {NoStop}%
\bibitem [{\citenamefont {Bertaut}(1962)}]{Bertaut_1962}%
  \BibitemOpen
  \bibfield  {author} {\bibinfo {author} {\bibfnamefont {E.~F.}\ \bibnamefont
  {Bertaut}},\ }\href {\doibase http://dx.doi.org/10.1063/1.1728635} {\bibfield
   {journal} {\bibinfo  {journal} {Journal of Applied Physics}\ }\textbf
  {\bibinfo {volume} {33}},\ \bibinfo {pages} {1138} (\bibinfo {year}
  {1962})}\BibitemShut {NoStop}%
\bibitem [{\citenamefont {Chapon}(2009)}]{Chapon_2009}%
  \BibitemOpen
  \bibfield  {author} {\bibinfo {author} {\bibfnamefont {L.~C.}\ \bibnamefont
  {Chapon}},\ }\href {\doibase 10.1103/PhysRevB.80.172405} {\bibfield
  {journal} {\bibinfo  {journal} {Phys. Rev. B}\ }\textbf {\bibinfo {volume}
  {80}},\ \bibinfo {pages} {172405} (\bibinfo {year} {2009})}\BibitemShut
  {NoStop}%
\bibitem [{\citenamefont {Reimers}\ \emph {et~al.}(1991)\citenamefont
  {Reimers}, \citenamefont {Berlinsky},\ and\ \citenamefont
  {Shi}}]{Reimers_1991}%
  \BibitemOpen
  \bibfield  {author} {\bibinfo {author} {\bibfnamefont {J.~N.}\ \bibnamefont
  {Reimers}}, \bibinfo {author} {\bibfnamefont {A.~J.}\ \bibnamefont
  {Berlinsky}}, \ and\ \bibinfo {author} {\bibfnamefont {A.-C.}\ \bibnamefont
  {Shi}},\ }\href {\doibase 10.1103/PhysRevB.43.865} {\bibfield  {journal}
  {\bibinfo  {journal} {Phys. Rev. B}\ }\textbf {\bibinfo {volume} {43}},\
  \bibinfo {pages} {865} (\bibinfo {year} {1991})}\BibitemShut {NoStop}%
\bibitem [{\citenamefont {Chernyshev}\ and\ \citenamefont
  {Zhitomirsky}(2015)}]{PhysRevB.92.144415}%
  \BibitemOpen
  \bibfield  {author} {\bibinfo {author} {\bibfnamefont {A.~L.}\ \bibnamefont
  {Chernyshev}}\ and\ \bibinfo {author} {\bibfnamefont {M.~E.}\ \bibnamefont
  {Zhitomirsky}},\ }\href {\doibase 10.1103/PhysRevB.92.144415} {\bibfield
  {journal} {\bibinfo  {journal} {Phys. Rev. B}\ }\textbf {\bibinfo {volume}
  {92}},\ \bibinfo {pages} {144415} (\bibinfo {year} {2015})}\BibitemShut
  {NoStop}%
\bibitem [{\citenamefont {Genz}\ and\ \citenamefont {Malik}(1980)}]{Genz1980}%
  \BibitemOpen
  \bibfield  {author} {\bibinfo {author} {\bibfnamefont {A.~C.}\ \bibnamefont
  {Genz}}\ and\ \bibinfo {author} {\bibfnamefont {A.}~\bibnamefont {Malik}},\
  }\href@noop {} {\bibfield  {journal} {\bibinfo  {journal} {J. Comput. Phys.
  and Appl. Maths.}\ }\textbf {\bibinfo {volume} {6}},\ \bibinfo {pages} {295}
  (\bibinfo {year} {1980})}\BibitemShut {NoStop}%
\bibitem [{\citenamefont {Berntsen}\ \emph {et~al.}(1991)\citenamefont
  {Berntsen}, \citenamefont {Espelid},\ and\ \citenamefont
  {Genz}}]{Berntsen1991}%
  \BibitemOpen
  \bibfield  {author} {\bibinfo {author} {\bibfnamefont {J.}~\bibnamefont
  {Berntsen}}, \bibinfo {author} {\bibfnamefont {T.~O.}\ \bibnamefont
  {Espelid}}, \ and\ \bibinfo {author} {\bibfnamefont {A.}~\bibnamefont
  {Genz}},\ }\href@noop {} {\bibfield  {journal} {\bibinfo  {journal} {ACM
  Transactions on Mathematical Software (TOMS)}\ }\textbf {\bibinfo {volume}
  {17}},\ \bibinfo {pages} {437} (\bibinfo {year} {1991})}\BibitemShut
  {NoStop}%
\bibitem [{\citenamefont {Igarashi}(1992)}]{Igarashi_1992}%
  \BibitemOpen
  \bibfield  {author} {\bibinfo {author} {\bibfnamefont {J.-I.}\ \bibnamefont
  {Igarashi}},\ }\href {\doibase 10.1103/PhysRevB.46.10763} {\bibfield
  {journal} {\bibinfo  {journal} {Phys. Rev. B}\ }\textbf {\bibinfo {volume}
  {46}},\ \bibinfo {pages} {10763} (\bibinfo {year} {1992})}\BibitemShut
  {NoStop}%
\end{thebibliography}

%

%======================================================================
% End
%======================================================================
\end{document}